\documentclass[aps,prb,twocolumn,showpacs,superscriptaddress,floatfix,
amsmath,amssymb,
]{revtex4-2}

\usepackage{graphicx}
\usepackage{dcolumn}
\usepackage{bm}
\usepackage{hyperref}
\usepackage[mathlines]{lineno}

\begin{document}

\title{\textit{First-principles} analysis of the interplay between electronic structure and volume change in colquiriite compounds during Li intercalation}

\author{A. F. Baumann}
 \email{aljoscha.baumann@fmf.uni-freiburg.de}
 \affiliation{University of Freiburg, Freiburg Materials Research Center (FMF), Stefan-Meier-Straße 21, 79104 Freiburg, Germany}
 
\author{D. Mutter}%
 \affiliation{Fraunhofer IWM, Wöhlerstraße 11, 79108 Freiburg, Germany}

\author{D. F. Urban}
\affiliation{University of Freiburg, Freiburg Materials Research Center (FMF), Stefan-Meier-Straße 21, 79104 Freiburg, Germany}
\affiliation{Fraunhofer IWM, Wöhlerstraße 11, 79108 Freiburg, Germany}

\author{C. Elsässer}
\affiliation{University of Freiburg, Freiburg Materials Research Center (FMF), Stefan-Meier-Straße 21, 79104 Freiburg, Germany}
\affiliation{Fraunhofer IWM, Wöhlerstraße 11, 79108 Freiburg, Germany}

\begin{abstract}
A main source of capacity fading in lithium-ion batteries is the degradation of the active cathode materials caused by the series of volume changes during charge and discharge cycles. The quaternary colquiriite-type fluorides Li$_x$CaFeF$\mathrm{_6}$ and Li$_x$CaCoF$\mathrm{_6}$ were reported to have negligible volume changes in specific Li concentration ranges, making the underlying colquiriite structure a promising candidate for so-called \textit{zero-strain} behavior. Using first-principles electronic structure calculations based on density functional theory with a Hubbard-$U$ correlation correction on the transition-metal ions, we systematically investigate the equilibrium volumes of the colquiriite-type fluorides Li$_x$CaMF$\mathrm{_6}$ with M =Ti, V, Cr, Mn, Fe, Co, and Ni at the Li concentrations $x$=0, 1, and 2. We elucidate the connection between the total volume of the structures and the local volumes of fluorine coordinated octahedra around the cations, and we find trends along the series of the \textit{3d} transition-metal elements. In the lithiation step from $x$=1 to $x$=2 we find volume changes of about 10~\%, and we discuss the discrepancy to the experimentally reported smaller value for Li$_x$CaFeF$\mathrm{_6}$. The suitability as cathode material was furhter investigated by calculating the theoretical voltages and capacities. From $x$=0 to $x$=1 we describe the compensating structural mechanisms that lead to an exceptionally small volume change of Li$_x$CaMnF$\mathrm{_6}$, which posseses a high theoretical voltage and moderate capacity. This compound is therefore a particularly promising zero-strain cathode material.
\end{abstract}

\maketitle


\section{Introduction}

Active cathode materials for lithium-ion batteries (LIBs) often suffer from mechanical degradation during charge and discharge cycles \cite{Edge_2021_degradation,Pender_2020_degradation,Han_2019_degradation,Hausbrand_2015_degradation}. Among other effects, phase transformations or changes in the lattice parameters can occur as a consequence of the intercalation and deintercalation of Li ions. The associated volume change varies from material to material. The widely used cathode material Li$_x$CoO$\mathrm{_2}$ (LCO), for example, undergoes a sequence of three distinct phase transitions during delithiation \cite{Reimers_1992_LCO_expansion}. The associated structural changes were found to be primarily responsible for the performance degradation of Graphite/LCO LIBs.\cite{Wang_2007} So-called \textit{zero-strain} (ZS) materials are characterized by very small volume changes during ionic charging and discharging and are therefore promising candidates for mechanically stable cycling. This is especially relevant in the context of all-solid-state LIBs, where the stability of the interfaces between the active electrode and the solid electrolyte particles is crucial \cite{koerver_2018,Wang_2023}.

Several ZS materials were reported in the literature and different mechanisms were proposed to explain the effect. Li$\mathrm{_4}$Ti$\mathrm{_5}$O$\mathrm{_{12}}$ is a well-known ZS anode material \cite{Zaghib_1998_LTO_ZS,Ohzuku_1995_LTO_ZS,Ziebarth_2014_LTO} for which the ZS behavior can be explained by the compensation of changing O-Ti-O bond angles and Li-O bond lengths during (dis-)charging \cite{Tian_2020_LTO_ZS}. Experimentally characterized metal-oxide ZS cathode materials for LIBs include the spinel-type compounds LiCoMnO$\mathrm{_4}$, where the ZS behavior is mainly ascribed to a small difference in the ionic radii of Co$\mathrm{^{3+}}$/Co$\mathrm{^{4+}}$ \cite{Ariyoshi_2018_ZS_LCOMNO4}, and Li$\mathrm{_2}$Ni$\mathrm{_{0.2}}$Co$\mathrm{_{1.8}}$O$\mathrm{_4}$, where the effects of contracting Co-O bonds and distortions of the oxygen sublattice cancel each other \cite{Ariyoshi_2019_ZS_LiNiCoO4}. In the disordered rocksalt-type Li-excess compounds Li$\mathrm{_{1.3}}$V$\mathrm{_{0.4}}$Nb$\mathrm{_{0.3}}$O$\mathrm{_2}$ and Li$\mathrm{_{1.25}}$V$\mathrm{_{0.55}}$Nb$\mathrm{_{0.2}}$O$\mathrm{_{1.9}}$F$\mathrm{_{0.1}}$, the ZS behavior is described by effects including the existence of transition-metal redox centers and electrochemical inactive elements, and a migration of Li from octahedral to tetrahedral sites  \cite{Zhao_2022}. 

In addition to oxide compounds, ZS behavior has also been observed for some fluoride compounds. In the tungsten-bronze-type compound K$\mathrm{_{0.6}}$FeF$\mathrm{_3}$ nearly-zero-strain sodiation was measured, whichwas attributed to the proper size of the cavities for Na ions in this open framework structure \cite{Han_2016_ZS_TB}. Koyama et al.\ first reported a compound crystallizing in the so-called \textit{colquiriite} structure in the context of ZS positive-electrode materials. With density functional theory (DFT) calculations in the local density approximation (LDA) without spin polarization, they obtained a volume change for Li$_x$CaCoF$\mathrm{_6}$ of 0.4~\% for a change of the Li concentration from $x$=1 to $x$=0. They explained the ZS behavior during delithiation by the reduction of Co-F bond lengths (due to the oxidation of Co) compensating the elongation of Li-F bond lengths (stronger F$\mathrm{^-}$-F$\mathrm{^-}$ repulsion due to the missing attraction between Li$\mathrm{^+}$ and F$\mathrm{^-}$ after the removal of Li) \cite{koyama_2000}. De Biasi and coworkers measured a small volume change below 0.5~\% in the colquiriite-type compound Li$_x$CaFeF$\mathrm{_6}$ in a Li concentration range of $1.0\leq x\leq 1.8$ \cite{de_biasi_licafef_2017}. This small volume expansion was attributed to the large absolute volume of the Li-F octahedron that mitigates the effect of the expanding Fe-F octahedron.

The colquiriite structure is named after the mineral LiCaAlF$\mathrm{_6}$, which was first found in Colquiri, Bolivia. Since then, many other compounds with this structure and the general formula LiA$\mathrm{^{2+}}$M$\mathrm{^{3+}}$F$\mathrm{_6}$ were successfully synthesized \cite{fleischer_zur_1982,Schaffers_1991,de_biasi_licafef_2017,Yin_1992,Yin_1993,Klimm1998NonstoichiometryOT,Ono_2001,Cadatal-Raduban_2021}. Most research focusing on colquiriites concerns their application as laser media. Especially Cr-doped LiCaAlF$\mathrm{_6}$ and LiSrAlF$\mathrm{_6}$ were investigated both experimentally and theoretically \cite{Payne_1988,Du_2012,LUONG201715,Shimizu_2017,DEMIRBAS_2019}.

Pawlak \textit{et al.} investigated the correlation between structural parameters of colquiriite structures by means of a statistical analysis of available experimental data \cite{PAWLAK_2001}. For a set of LiCaMF$\mathrm{_6}$ structures they showed that different M ions have a strong influence on both lattice parameters $a$ and $c$  as well as on the Li-F bond length: With larger M-F bond lengths the Li(2c)-F bond length increases, too. However, no structural correlation was detected between the M-F and Ca-F bond lengths.

Motivated by the stability of the crystalline colquiriite framework for various element combinations on the cation sublattices, and by the reports of ZS behavior for representatives of this structure type, we apply electronic structure calculations to systematically analyze the volume changes of the colquiriite compounds Li$_x$CaMF$\mathrm{_6}$ for the lithiation steps from $x$=0 to $x$=1 and from $x$=1 to $x$=2, considering the series of \textit{3d} transition-metal elements (TMs) Ti, Cr, V, Mn, Fe, Co, and Ni on the M site. We investigate different magnetic configurations of the TM \textit{3d} electrons since these influence the total energy and volume of the compounds. Electronic structure calculations are a valuable tool to study volume effects related to magnetism in metals. Previous reports that relate the electronic structure to volume effects include, e.g., the so-called Invar effect \cite{Schilfgaarde_1999_Invar_effect}, the relationship between magnetism and volume of cubic Fe, Co, and Ni phases \cite{Moruzzi_1986_volume_effect} and magneto-volume effects in iron and cobalt nitrides \cite{Houari_2010_magneto_volume}. 

The paper is organized as follows: In Sec.\ \ref{sec:methods_structure}, the colquiriite structure is described, followed by a description of the computational details in Sec.\ \ref{sec:methods_compdetails}. The results of the atomistic DFT calculations for the determination of the magnetic ground states are presented in Section \ref{sec:results_groundstate} and the volume changes between those states for different Li concentrations are presented in Sec.\ \ref{sec:results_volchanges}. The results are discussed in Sec.\ \ref{sec:discussion}: first we compare ionic radii for different $U$ values (\ref{discussion:U_value_choosing}), then we discuss the relationship between the changes in ionic radii (viz. volumes of octahedra) during lithiation and the electronic structure of the TM element (\ref{sec:discussion_ionic_radii_curves}), followed by an explanation of how the local volumes of octahedra of fluorine anions determine the total unit-cell volume of the crystal structure (\ref{sec:discussion_global_volume_change}). Our conclusions are given in Section \ref{sec:conclusion}.


\section{Theoretical approach} \label{sec:methods}

\subsection{Colquiriite structure}\label{sec:methods_structure}
\begin{figure}
    \centering
    \includegraphics[width=0.75\columnwidth]{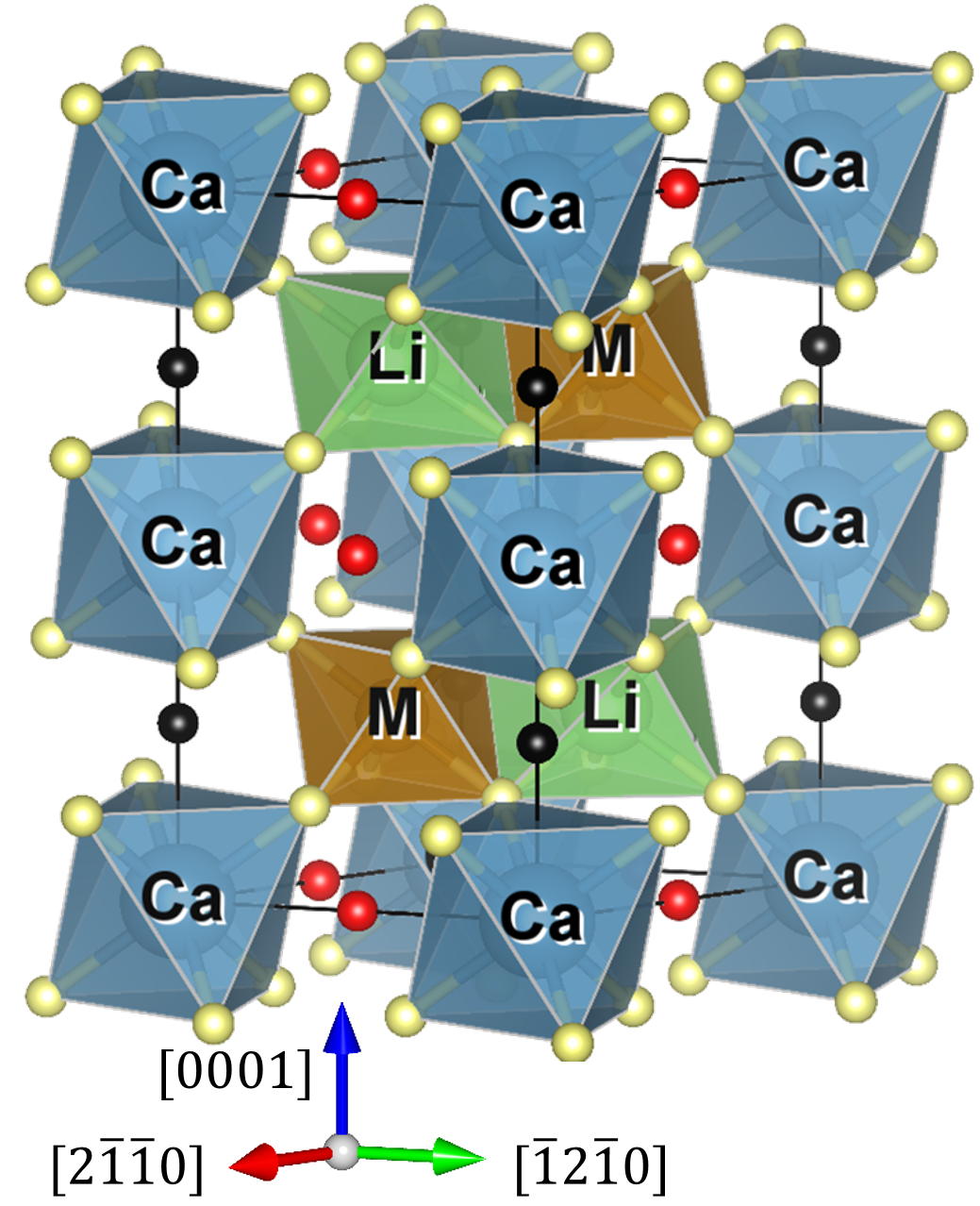}
    \caption{Trigonal unit cell of the colquiriite-type structure of LiCaMF$\mathrm{_6}$ (M=Ti, V, Cr, Mn, Fe, Co, and Ni). The octahedra of M cations coordinated by fluorine anions (yellow spheres) are labelled by the respective central cation M. Black and red spheres at Wyckoff positions 2a and 6g, respectively, indicate the possible sites for additional Li ions.}
    \label{fig:licammf6_structure}
\end{figure}

Colquiriite-type compounds crystallize in the trigonal space group $P\Bar{3}1c$. The unit cell of a LiCaMF$\mathrm{_6}$ crystal consists of two formula units (f.u.), i.e., 18 atoms, displayed in Figure \ref{fig:licammf6_structure}. The lattice parameter $a$ is defined along the $[2\Bar{1}\Bar{1}0]$ direction (or equivalently along the $[\Bar{1}2\Bar{1}0]$ direction) and $c$ is defined along the $[0001]$ direction. Along the $[0001]$ direction, the crystal consists of alternating layers of fluorine coordinated Ca and Li/M octahedra. The Ca and M ions occupy the Wyckoff positions (WPs) 2b and 2d, respectively. If there is one Li atom per f.u.\ it is located at the WP 2c. Additional possible sites for Li ions in case of higher Li concentrations are the WPs 2a and 6g (fractional coordinates: $[0,0,1/4]$ and $[0,1/2,1/2]$, where for the latter, three inequivalent arrangements of Li ions on these sites are possible). We describe the energetic hierarchy of the different Li occupations in more detail in Sec.\ \ref{sec:results:Li_wyckoff}. Table \ref{tab:wyckoff_positions} lists the coordinates of the symmetric sites of the colquiriite structure.

\begin{ruledtabular}
\begin{table}
\centering
\begin{tabular}{ cccccc} 
 Atom & Multiplicity & Wyckoff letter & x & y & z \\
\hline
Ca  & 2 & b & 0 & 0 & 0\\ 
Li & 2 & c & 1/3 & 2/3 & 1/4\\
M  & 2 & d & 2/3 & 1/3 & 1/4\\ 
F  & 12 & i & 0.376 & 0.032 & 0.144\\ 
\end{tabular}
\caption{Wyckoff positions and fractional coordinates of the asymmetric unit of the colquiriite structure (space group $P\Bar{3}1c$). The fluorine atoms, which occupy the general position $[x,y,z]$ are here exemplarily given for LiCaAlF$\mathrm{_6}$ according to Ref.\ \cite{Viebahn_1971_Colquiriite_structure}.}
\label{tab:wyckoff_positions}
\end{table}
\end{ruledtabular}

\subsection{Computational details}\label{sec:methods_compdetails}
We determine the equilibrium volumes $V_0$ of the colquiriite unit cells by fitting the Murnaghan equation of state (EOS) \cite{Murnaghan_EOS_1944} to a dataset of minimal total energies calculated for different unit-cell volumes. At each given volume we optimize the unit-cell shape and the atomic coordinates. 

The procedure of using energy-volume (EV) curves \cite{Silvi_EOS,TYUTEREV_EOS_2006} gives more accurate results for $V_0$ compared to the volume optimization by stress minimization, with routines implemented in most atomistic simulation codes. It avoids numerical errors in the Pulay stresses that originate from the dependence of the incompleteness of the plane-waves basis on the unit-cell volume. In addition, by means of the EV curves one can clearly distinguish minimum energies and equilibrium volumes of electronically or magnetically stable (ground-state) and metastable structures as long as those follow separate EV curves, rather than only getting one – global or local – energy-minimum point as in the case of unconstrained volume optimization by stress minimization.

All DFT calculations were carried out with the Vienna ab initio simulation package (VASP) \cite{vasp_general}. The Perdew-Burke-Ernzerhof generalized gradient approximation (PBE-GGA)\cite{perdew_generalized_1996} was used for the exchange-correlation functional. Interactions between valence electrons and ionic cores are treated with the projector-augmented wave (PAW) method \cite{vasp_paw}: For Li, the \textit{1s}, \textit{2s} and \textit{2p} orbitals, for Ca, the \textit{3s}, \textit{3p} and \textit{4s} orbitals, for the considered TMs, the \textit{3s}, \textit{3p}, \textit{3d} and \textit{4s} orbitals, and for F, the \textit{2s} and \textit{2p} orbitals are taken as valence states.

Similar to the situation in oxides, the localized electrons of the \textit{3d} TMs in fluorides are often not well described by DFT with LDA or GGA approximations. A Hubbard-$U$ correlation-correction term can mitigate the respective self-interaction error which leads to a too weak localization of the \textit{3d} electrons \cite{Anisimov_1991_DFTU,Tolba18_DFTU,Himmetoglu_2013_DFTUreview,Correa_2018_DFTU,Kang_2022_DFTU,Tada_2021_DFTU,dilucente2023crossover_DFTU}. The value of the $U$ parameter is often chosen semi-empirically, e.g., by tuning it to match the calculated electronic band gap with experimentally measured values. However, the scarce experimental data for the electronic structure of colquiriite-type fluoride compounds, especially for the band gap, does not allow a corresponding adjustment of the $U$ value for our study. We therefore pursue a more generic comparison between theoretically calculated and experimentally measured quantities: we compare the change in the ionic radii of the electrochemically active cations during lithiation, extracted from the ground state structures for different $U$ values, to the change in the ionic radii given by Shannon \cite{Shannon_1976_ionic_radii}. $U$ is chosen in such a way that our calculated values and the Shannon radii match best for the considered series of \textit{3d} TM ions. We followed the DFT+$U$ approach of Dudarev \textit{et al.} \cite{dudarev_electron-energy-loss_1998} and compared calculations with values of $U\mathrm{_{eff}}$=0 (no $U$ correction), 4 eV, and 8 eV on the\textit{3d} orbitals of the M ions, where $U\mathrm{_{eff}} = U-J$. Here, $U$ denotes the on-site Coulomb term and $J$ the site exchange term. To simplify the notation, $U\mathrm{_{eff}}$ will be denoted just as $U$ in the following.

We set the energy cutoff for the plane-waves basis to 700 eV and use a 6$\times$6$\times$3 Monkhorst-Pack \cite{monkhorst} \textit{k}-point mesh with a Gaussian smearing of 0.05 eV \cite{Fu_1983_smearing,Elsaesser_1994_smearing} for the Brillouin-zone integrations. We converged the total energies to 5$\times$10$^{-6}$~eV in the electronic self-consistency loop and the interatomic forces to 0.05 eV/{\AA} in the ionic relaxation loop.


\section{Results} \label{sec:results}

\subsection{Ground state configurations}\label{sec:results_groundstate}

In order to find the state of lowest energy (ground state) for each considered compound Li$_x$CaMF$\mathrm{_6}$, we investigated the different possible sites for Li ions in the colquiriite structure and the different possible local magnetic moments of the M ions. We then calculated the relative energetic ordering of those compounds.

\subsubsection{Occupation of Li sites}\label{sec:results:Li_wyckoff}

To figure out which of the WPs 2a, 2c, and 6g are favored to be occupied by Li for the Li concentrations $x$=1 and $x$=2, we set up the corresponding unit cells and calculated the EV curves. In addition, we deliberately displaced the Li atoms slightly from their high-symmetry WPs, which however always resulted in a relaxation back to the WPs. For each crystal composition with TM element M and Li concentration $x$, the energies of the structures with different Li occupations were calculated for the three considered $U$ values, 0 eV, 4 eV and 8 eV, and for the possible variety of magnetic states (i.e., high-spin and low-spin states, which are described in detail in Sec.\ \ref{sec:results_magnetic_states}). In the first lithiation step from $x$=0 to $x$=1, we found that the 2c sites are always the energetically favored positions for Li by at least 0.28 eV/f.u.

For the second lithiation step from $x$=1 to $x$=2, we accordingly considered unit cells with Li on the 2c sites and placed the two additional Li atoms on the 2a or 6g sites. It turned out that an occupation of the 2a sites leads to structures that are lower in energy by at least 0.11 eV/f.u. than the structures where the 6g sites are occupied.

\subsubsection{Local magnetic moments of TM ions}\label{sec:results_magnetic_states}

\begin{figure}
    \centering
    \includegraphics[width=\columnwidth]{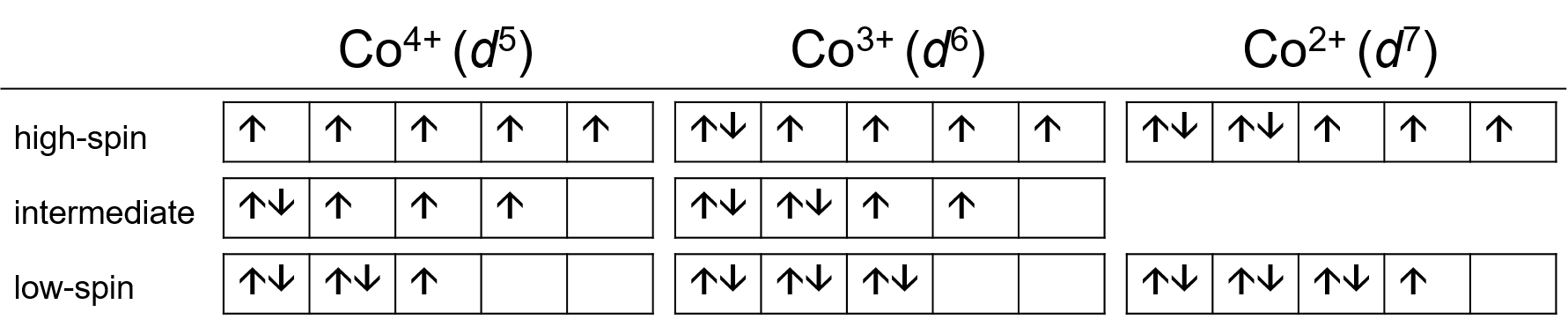}
    \caption{Possibilities to distribute the \textit{3d} electrons of Co in the \textit{3d} orbitals for three different oxidation states. The arrows indicate the two different spin directions of the electrons. The corresponding local magnetic moments of the ions are obtained by counting the unpaired electrons.}
    \label{fig:scheme_mag_moms}
\end{figure}

Following the idealized chemical notion of occupying atomic orbitals with electrons/spins, there are several distinct possibilities to formally distribute the valence electrons of a TM ion on the atomic \textit{3d}  orbitals, which lead to different local magnetic moments. Figure \ref{fig:scheme_mag_moms} exemplary illustrates this for the Co ion in the three oxidation states, Co$\mathrm{^{4+}}$, Co$\mathrm{^{3+}}$, and Co$\mathrm{^{2+}}$. Without considering the corresponding energies, one can in principle distinguish a \textit{high-spin} (HS) state with a maximum number of unpaired electrons, a \textit{low-spin} (LS) state with a minimum number of unpaired electrons, and a state with an intermediate spin configuration. Co has the valence electron configuration \textit{3d}$\mathrm{^{7}}$\textit{4s}$\mathrm{^{2}}$. With the \textit{4s} electrons taken first by fluorine anions, this leads to magnetic moments of 5 $\mu\mathrm{_B}$ (HS) and 1 $\mu\mathrm{_B}$ (LS) for Co$\mathrm{^{4+}}$, 4 $\mu\mathrm{_B}$ (HS) and 0 $\mu\mathrm{_B}$ (LS) for Co$\mathrm{^{3+}}$, and 3 $\mu\mathrm{_B}$ (HS) and 1 $\mu\mathrm{_B}$ (LS) for Co$\mathrm{^{2+}}$, as sketched in Fig.\ \ref{fig:scheme_mag_moms}.

The three oxidation states 4+, 3+, and 2+ will be considered in the following for each M cation, since they are formally adopted by those ions in the colquiriite structures with Li concentrations $x$=0, 1, and 2, respectively, as a consequence of total charge neutrality.

For all compounds, we carried out a series of DFT calculations by initializing the magnetic moments at all integer values between 0 $\mu\mathrm{_B}$ and 5 $\mu\mathrm{_B}$. After the structural relaxations, at most two states were found to be stable, namely one HS and one LS state. No initially prepared intermediate state could be stabilized finally. For Ti$\mathrm{^{4+}}$, there is no \textit{3d} electron, and for Ti$\mathrm{^{3+}}$ and V$\mathrm{^{4+}}$, there is only one \textit{3d} electron and hence only a single-spin state. For some cations with more than one \textit{3d} electron, i.e., where formally more than one spin state may be expected, only one spin state was found to be stable (e.g., a HS state of V$\mathrm{^{2+}}$ with three unpaired \textit{3d} electrons and 3 $\mu\mathrm{_B}$).

Since there are two M atoms in the unit cell of the colquiriite structure considered in this work, one can distinguish two collinear arrangements of magnetic moments, a ferromagnetic (FM) one, where the local magnetic moments on the two M ions are oriented parallel to each other, and an antiferromagnetic one (AFM), where the local magnetic moments are oriented anti-parallel. For the example of Li$_x$CaCoF$\mathrm{_6}$, we set up the two arrangements with different initial magnetic moments and performed total energy calculations for structures with $x$=0, 1, and 2, and $U$=0 eV, $4$ eV, and $8$ eV. The local magnetic moments of the HS and LS states on the individual M ions were found to be identical for both FM and AFM arrangements, with differences in energy between those of less than 0.0074 eV/f.u. Similar results for other exemplarily tested compounds indicate negligibly weak exchange interactions between the local magnetic moments on the different M ions. Therefore, we present only the results for FM arrangements in the following.

\begin{figure}
    \centering
    \includegraphics[width=\columnwidth]{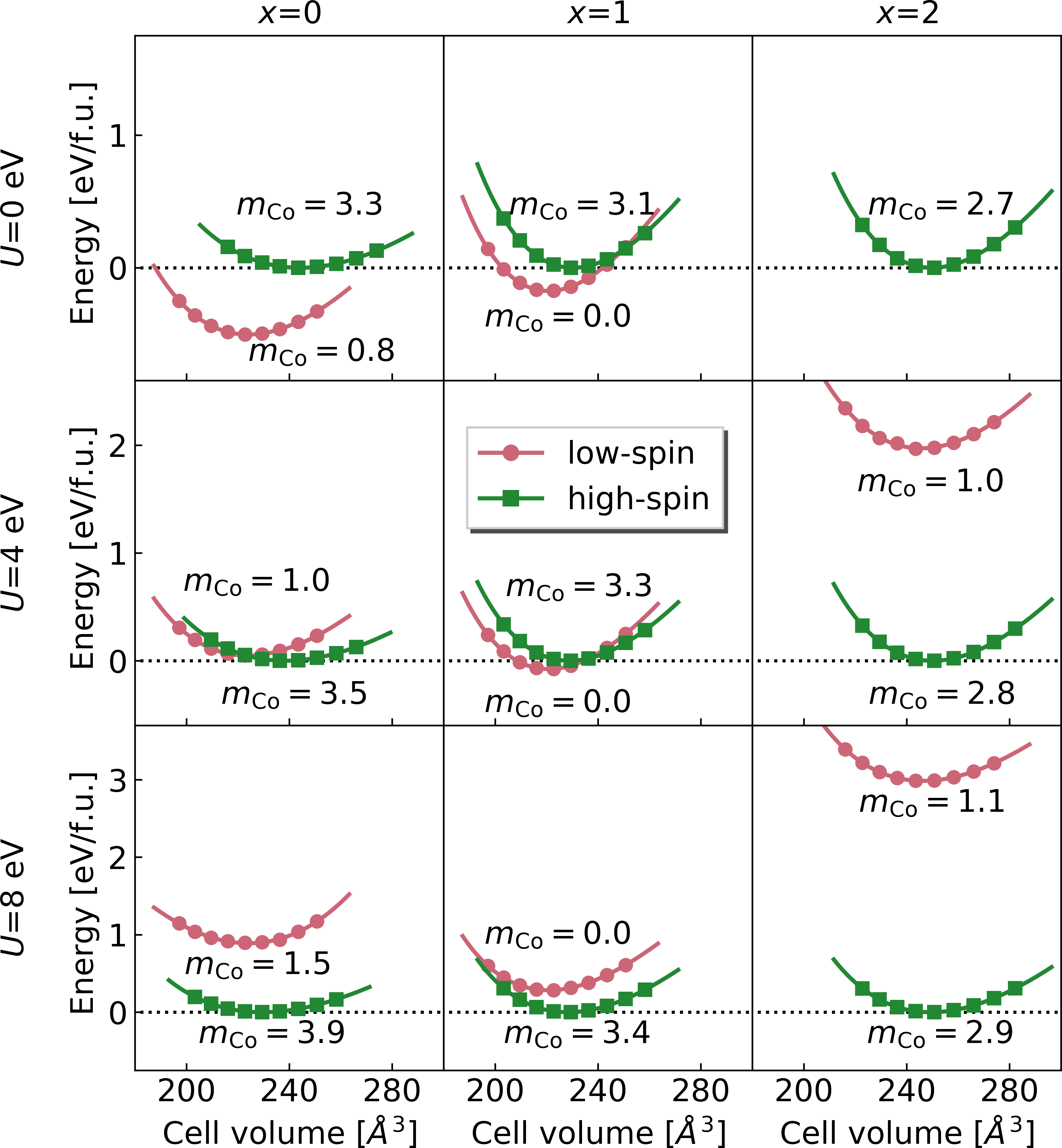}
    \caption{EV curves of Li$_x$CaCoF$\mathrm{_6}$ for the three Li concentrations ($x$=0, 1, and 2), and for the three $U$ parameters ($U$=0 eV, $4$ eV, and $8$ eV). In each of the nine panels, the energy is referenced to the energy minimum of the high-spin (HS) EV curve. Filled circles mark calculated total energies per f.u. obtained at fixed volumes and variable cell shape, and solid lines indicate the fitted EOS. The magnetic moments (in units of $\mu\mathrm{_B}$) of the Co ions after relaxation are given at each curve.}
    \label{fig:licacof_envol_front}
\end{figure}

\begin{ruledtabular}
\begin{table*}
\centering
\begin{tabular}{lp{1.62cm}p{1.62cm}p{1.62cm}p{1.62cm}p{1.62cm}p{1.62cm}p{1.62cm}}
& Ti & V   & Cr & Mn & Fe  & Co  & Ni  \\ \hline
$x$=0; M$^\mathrm{4+}$\\
$U$=0 &  &  & HS (-0.95) & HS (*)  & HS (-0.75) & LS (0.5) & LS (*)\\
$U$=4 &  &  & HS (-2.5) & HS (*) & HS (-1.99) & HS (-0.05) & LS (1.0)\\
$U$=8 &  &  & HS (*)    & HS (*) & HS (-3.53) & HS (-0.89) & LS (0.13)\\\\

$x$=1; M$^\mathrm{3+}$  \\ 
$U$=0 &  & HS (-0.99) & HS (*) & HS (-1.66) & HS (-1.11) & LS (0.17) & LS (*)\\
$U$=4 &  & HS (*) & HS (*) & HS (-2.43) & HS (-1.96) & LS (0.08) & HS (-0.26)\\
$U$=8 &  & HS (*) & HS (*) & HS (-2.73) & HS (*) & HS (-0.28) & HS (-0.48)\\\\

$x$=2; M$^\mathrm{2+}$  \\
$U$=0 & HS (*) & HS (*) & HS (-1.59) & HS (-2.33) & HS (-1.4) & HS (*) & HS (-1.17)\\
$U$=4 & HS (*) & HS (*) & HS (-2.2) & HS (*) & HS (-1.81) & HS (-1.97) & HS (*)\\
$U$=8 & HS (*) & HS (*) & HS (-2.71) & HS (*) & HS (-2.06) & HS (-2.98) & HS (*)\\
\end{tabular}
\caption{Magnetic ground states for the compounds Li$_x$CaMF$\mathrm{_6}$ with M = Ti, V, Cr, Mn, Fe, Co, and Ni, for $x$=0, 1 and 2 (corresponding to the formal oxidation states of the M cations 4+, 3+, and 2+) and for $U$=0 eV, $4$ eV and $8$ eV applied to the M ions. In parenthesis $\Delta E\mathrm{^{HL}_0}$ is given. (*) indicates that only the magnetic state presented in the Table was stable after relaxation. All numerical values are given in eV.}
\label{tab:delta_energies}
\end{table*}
\end{ruledtabular}

Figure \ref{fig:licacof_envol_front} displays the calculated energy versus volume data and the fitted EV curves for $x$=0, 1, and 2, and for $U$=0 eV, $4$ eV and $8$ eV for Li$_x$CaCoF$\mathrm{_6}$. The results for HS and LS states are displayed in green and red color, respectively. For Li$\mathrm{_2}$CaCoF$\mathrm{_6}$ calculated with $U$=0 eV, only one magnetic state was stable after relaxation. For the other cases, the energy ordering between HS and LS states depends on the value of $U$ as well as on the Li concentration $x$. As described above, the additional Li reduces the TM cation which results in different distributions of the \textit{3d} electrons with different energies. The magnetic moments increase in magnitude with higher values of $U$ as expected since $U$ forces the electrons to their respective M sites. For the HS states, the calculated values of the magnetic moments are less than the formally expected integer numbers (5 $\mu\mathrm{_B}$, 4 $\mu\mathrm{_B}$, and 3 $\mu\mathrm{_B}$ for $x$=0, 1 and 2, respectively), but approach them for increasing $x$. The magnetic moments of the M ions in the LS states correspond well to the formal oxidation numbers (0, 1, 0 for $x$=0, 1, 2). The magnetic moments of all TM ions are reported in the Supplemental Material \cite{Supplemental_material}.\\
We extract two quantities from each EV curve: (i) the equilibrium energy, $E\mathrm{_0}$, from which we can calculated $\Delta E\mathrm{^{HL}_0}=E\mathrm{^{HS}_0}-E\mathrm{^{LS}_0}$, the difference in minimum energy between the HS and LS state, for each combination of $x$ and $U$ and (ii) the equilibrium volumes $V\mathrm{_0}$ of the ground states at the different Li concentrations $x$. From those volumes we can calculate $\Delta V\mathrm{_0}$, the volume change of the unit cell due to the Li addition.

Table \ref{tab:delta_energies} summarizes the magnetic ground states for all the studied Li$_x$CaMF$\mathrm{_6}$ compounds and different $U$ values together with $\Delta E\mathrm{^{HL}_0}$, where applicable. For M=Ti at $x$=0 (Ti$\mathrm{^{4+}}$) and $x$=1 (Ti$\mathrm{^{3+}}$) and for M=V at $x$=1 (V$\mathrm{^{4+}}$), only a zero-spin or a single-spin state is possible, as explained above and therefore no distinction between LS or HS is meaningful.

For most of the compositions, the HS state is preferred. The LS state was only found to be more stable in some cases containing Co$\mathrm{^{3+}}$/Co$\mathrm{^{4+}}$ or Ni$\mathrm{^{3+}}$/Ni$\mathrm{^{4+}}$ and with specific values of $U$.\\
Our findings for TM fluorides are in good agreement with the findings for TM oxides reported by Jia \textit{et al.} \cite{3d_TM_oxide_spin}, who have analyzed the distribution of magnetic moments of the 3$d$ TM ions, which were calculted by first-principles for a large set of compounds from the Materials Project database \cite{jain_commentary_2013_MP}. For Ti, V, Cr, Mn and Fe the magnetic moments corresponding to the HS states were the ones occurring most often. For some oxidation states of Co and Ni no clear preferred spin state was found. In the fluoride compounds calculated in this work, for Co$\mathrm{^{4+}}$ and the $U$ values $U$=4 eV and $U$=8 eV the HS spin state was the energetically favorable one, whereas this state was not found for oxides by Jia \textit{et al.} \cite{3d_TM_oxide_spin}. They found for Ni$\mathrm{^{3+}}$ the low-spin and intermediate-spin state most often, which goes well in line with the experimental findings, that  Ni$\mathrm{^{3+}}$ in LiNiO$\mathrm{_{2}}$ is considered to be in the LS state \cite{LNO_review}.  Other DFT+$U$ studies with a wide range of applied $U$ values (2.45 eV \cite{Tsebesebe_2022_LNO}, 5.96 eV \cite{chakraborty_predicting_2018}, 3 eV, 6 eV and 9 eV \cite{saritas_charge_2020_LNO}) agree with that. In the colquiriite fluorides calculated in this work, we found the LS state of Ni$\mathrm{^{3+}}$ only for $U$=0 and otherwise the HS state. Therefore in general, the preferred spin state depends strongly on the oxidation state of the TM ions and only for some cases the magnetic ground state is influenced by the type of ligand surrounding the TM ion or the applied $U$ value. A trend can be observed that the HS states become more stable with higher $U$ values. This was reported earlier for Fe$\mathrm{^{2+}}$ compounds by Mariano \textit{et al.} \cite{Mariano_2020}, who calculated the energy differences between HS and LS states for six molecular complexes with different ligands applying DFT+$U$ with $U$ values between 0 eV and 8 eV.

\subsection{Local and global volume changes}\label{sec:results_volchanges}

From the obtained crystal structures of the $U$-dependent magnetic ground states we now analyze the ionic radii of the M cations as function of the Li concentration. We compare the results to Shannon’s empirical set of ionic radii extracted from experimental mineral-structure data. This comparison enables us to select a value for $U$, which fits best to experimental structure data. This procedure is discussed in more detail in Sec.\ \ref{discussion:U_value_choosing}. Overall, we found that $U$=4 eV gives the best agreement (within the values examined in this work) of our and Shannon’s radii for both lithiation steps. Therefore, we use this particular $U$ value with good confidence for all the calculations described in the following.

\begin{figure}
    \centering
    \includegraphics[width=0.9\columnwidth]{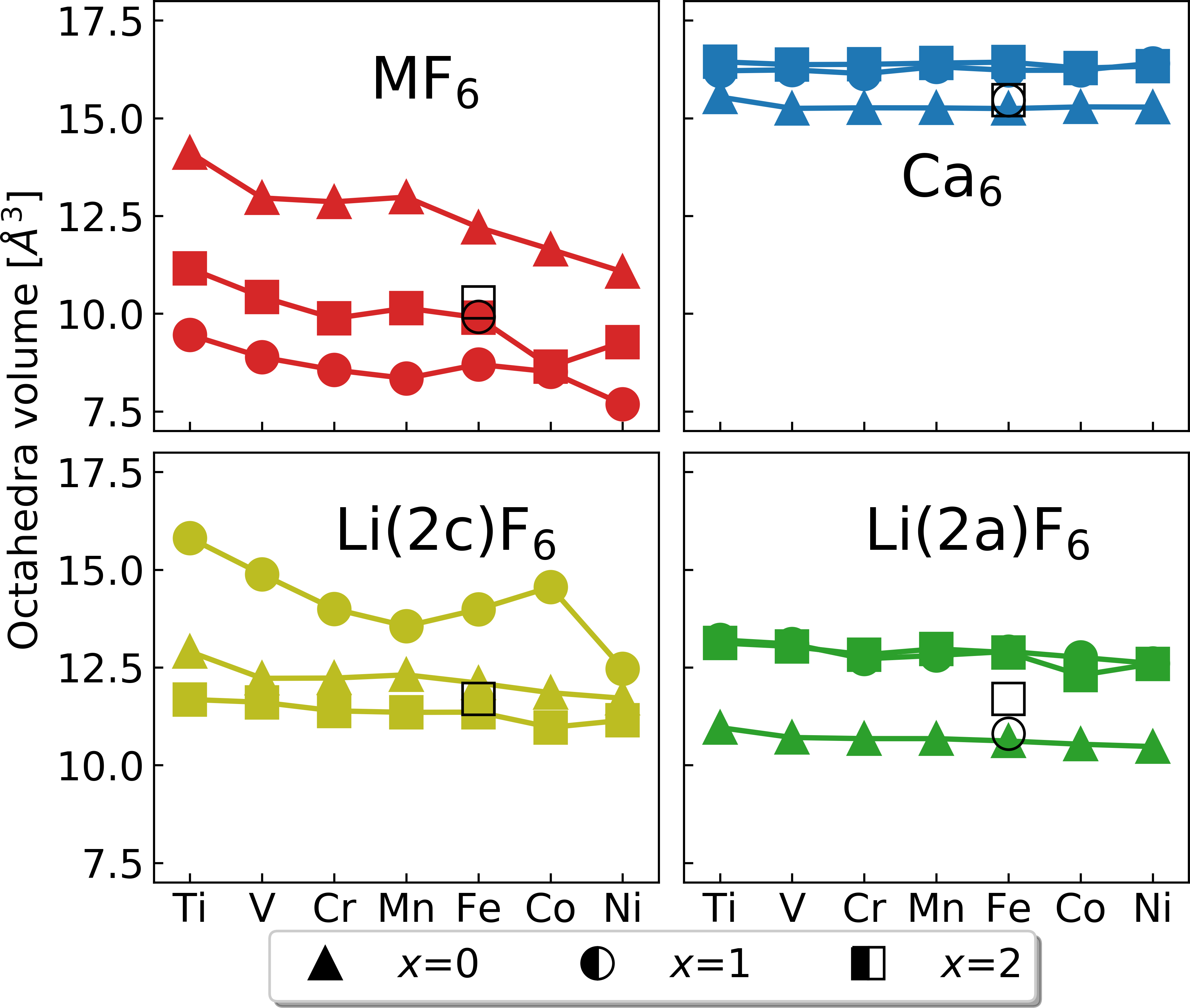}
    \caption{Absolute volumes of the MF$\mathrm{_6}$, LiF$\mathrm{_6}$, and CaF$\mathrm{_6}$ octahedra for the magnetic ground state structures at $U$=4 eV. The experimental results by de Biasi \textit{et al.} \cite{de_biasi_licafef_2017} for Li$_x$CaFeF$\mathrm{_6}$ are shown with open symbols.  The scale of the y axis is the same for all the plots. Lines between data points serve as guide for the eye.}
    \label{fig:absolute_octahedra_volumes}
\end{figure}

Figure \ref{fig:absolute_octahedra_volumes} depicts the absolute volumes of fluorine octahedra around the M, Li and Ca cations. Along the \textit{3d} series the MF$\mathrm{_6}$ octahedron volume decreases, which can be explained by the increasing nuclear charge of the atoms. The calculated octahedra volumes follow qualitatively similar trends as the experimentally measured bond lengths reported by Pawlak \textit{et al.} \cite{PAWLAK_2001} for colquiriite compounds of the type LiCaMF$\mathrm{_6}$: together with a lower MF$\mathrm{_6}$ octahedron volume (lower M-F bond length) the LiF$\mathrm{_6}$ (2c) octahedron volume (Li-F bond length) also decreases and the CaF$\mathrm{_6}$ octahedra volumes are not correlated with the M-F bond lengths.

The experimental results from de Biasi \textit{et al.} \cite{de_biasi_licafef_2017} for the octahedra volumes are shown for the lithiation concentrations $x$=1 and $x$=2. It is important to note, that the maximal concentration that was achieved in the experiments was $x$=1.79. The $\mathrm{^{57}}$Fe Mössbauer spectroscopy at that concentration however showed, that most of the iron ions were in the oxidation state Fe$\mathrm{^{2+}}$, i.e., similar to the oxidation state of the iron ions in the $x$=2 calculations of this work. In the experimental work only the average value of the LiF$\mathrm{_6}$ octahedra at $x$=1.79 was reported.\\
The value for the CaF$\mathrm{_6}$ octahedra volume measured experimentally is closer to the value obtained in the calculations at the lithium concentration $x$=2 than at $x$=1. De Biasi \textit{et al.} reported no change in the volume of theseoctahedra. The values for the FeF$\mathrm{_6}$ and LiF$\mathrm{_6}$ (2c) octahedra volumes at $x$=1 show a good agreement between the calculations of this work and the experimental results. For the higher Li concentration however, especially the volume of the FeF$\mathrm{_6}$ octahedron volume was measured to be much lower than the calculated one.

\begin{figure}
    \centering
    \includegraphics[width=\columnwidth]{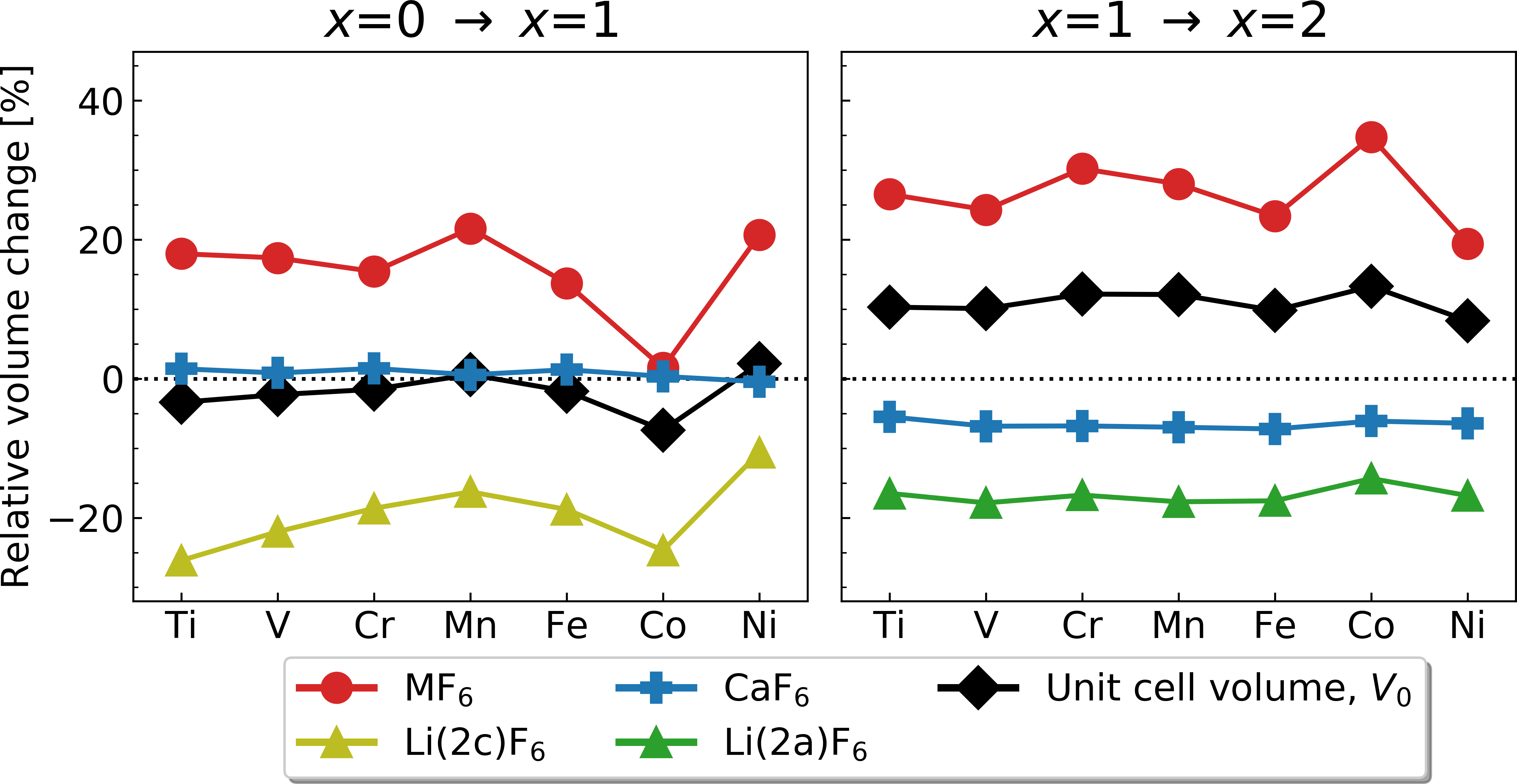}
    \caption{Relative volume changes of the MF$\mathrm{_6}$, LiF$\mathrm{_6}$, and CaF$\mathrm{_6}$ octahedra, for the ground states obtained for $U$=4 eV. For clarity only the changes of the Li octahedra, which are occupied in the corresponding lithiation step are shown. Lines between data points serve as guide for the eye.}
    \label{fig:U4_all_octahedra_CellVolume}
\end{figure}

In Figure \ref{fig:U4_all_octahedra_CellVolume} the relative volume changes of the fluorine octahedra in the two lithiation steps are displayed. 
For all the considered \textit{3d} TM elements M, the volumes of the octahedra around the Li sites, that become occupied during the two lithiation steps, get smaller. This is because of a reduction of the repulsion of the negatively charged fluorine anions of a F$\mathrm{_6}$ octahedron, when a positively charged Li ion occupies its center. For the step from $x$=0 to $x$=1, the volume changes of the LiF$\mathrm{_6}$ (2c) octahedra behave qualitatively similar along the \textit{3d} series as the volume changes of the MF$\mathrm{_6}$ octahedra, which is due to shared F-F edges between them (\textit{cf.} Figure \ref{fig:licammf6_structure}). In the second lithiation step the LiF$\mathrm{_6}$ (2a) octahedra become occupied, which also share F-F edges with the MF$\mathrm{_6}$ octahedra. However, they also share their top and bottom faces (in $[0001]$ direction) with two CaF$\mathrm{_6}$ octahedra. This leads to a less pronounced correlation between MF$\mathrm{_6}$ and LiF$\mathrm{_6}$ (2a) octahedra.

The CaF$\mathrm{_6}$ octahedra hardly change their volume in the first lithiation step,  independent of the element M. However, in the second lithiation step, there is a nearly constant contraction by about 5-7~\% of the volumes of CaF$\mathrm{_6}$ octahedra for all the considered M ions.

The relative change of the unit-cell volume of the colquiriite crystal, $\Delta V_0^{\rm (rel.)}  = (V_0[x+1]-V_0[x])/V_0[x]$, is also displayed in Fig.\ \ref{fig:U4_all_octahedra_CellVolume}.
In the first lithiation step (from $x$=0 to $x$=1), $\Delta V_0^{\rm (rel.)}$ adopts negative values between 0 and -7~\% for most of the considered compounds, or values slightly above zero for M=Mn and M=Ni. Especially for M=Mn, a remarkably small volume change  of 0.5~\%  indicates indeed a ZS behavior. For most of the compounds, volumes are compared between ground states, which have the same spin state (mainly HS, \textit{cf.}\ Tab.\ \ref{tab:delta_energies}). Only for Li$_x$CaCoF$\mathrm{_6}$ (HS to LS) and Li$_x$CaNiF$\mathrm{_6}$ (LS to HS), there is a concomitant change of the magnetic ground state. For M=Co, this transition causes the pronounced deviation of the volume change to a negative value of $-7.3\ \%$, more negative than the values of the other systems, in which no such transition between magnetic states takes place. For M=Ni, the different magnetic ground states do not lead to such a pronounced effect.
 
In the second lithiation step (from $x$=1 to $x$=2) all of the volume changes of the unit cells are positive within a range of about 8.5 - 12~\%. Only the compound containing Co exhibits a LS to HS transition, which results in a slightly higher value of 13.3~\%, compared to the other M cations. The large relative change of the CoF$\mathrm{_6}$ octahedra in this lithiation step stems from its volume at $x$=1 being small compared to the volumes of the other MF$\mathrm{_6}$ octahedra, while the absolute change of the octahedra volume is comparable to others (\textit{cf.}\ Fig.\ \ref{fig:absolute_octahedra_volumes}).

The increasing volumes of the MF$\mathrm{_6}$ octahedra affect the unit cell volumes, which can be seen from the qualitatively similar shapes of the red and black curves. The volume expansion of the MF$\mathrm{_6}$ octahedra by up to 20~\% must be, at least partially, compensated by other effects to obtain the overall negative or very small positive changes of the unit-cell volumes. The joint effect of the local volume changes of the different octahedra on the total unit-cell volume of the colquiriite  structure will be discussed in Section \ref{sec:discussion_global_volume_change}. The absolute volumes ($V_0$) of the unit cells for all compounds and $U$ values are listed in the Supplemental Material \cite{Supplemental_material}.

\subsection{Theoretical capacities and voltages}\label{sec:capacities_voltages}

\begin{figure}
    \centering
    \includegraphics[width=0.7\columnwidth]{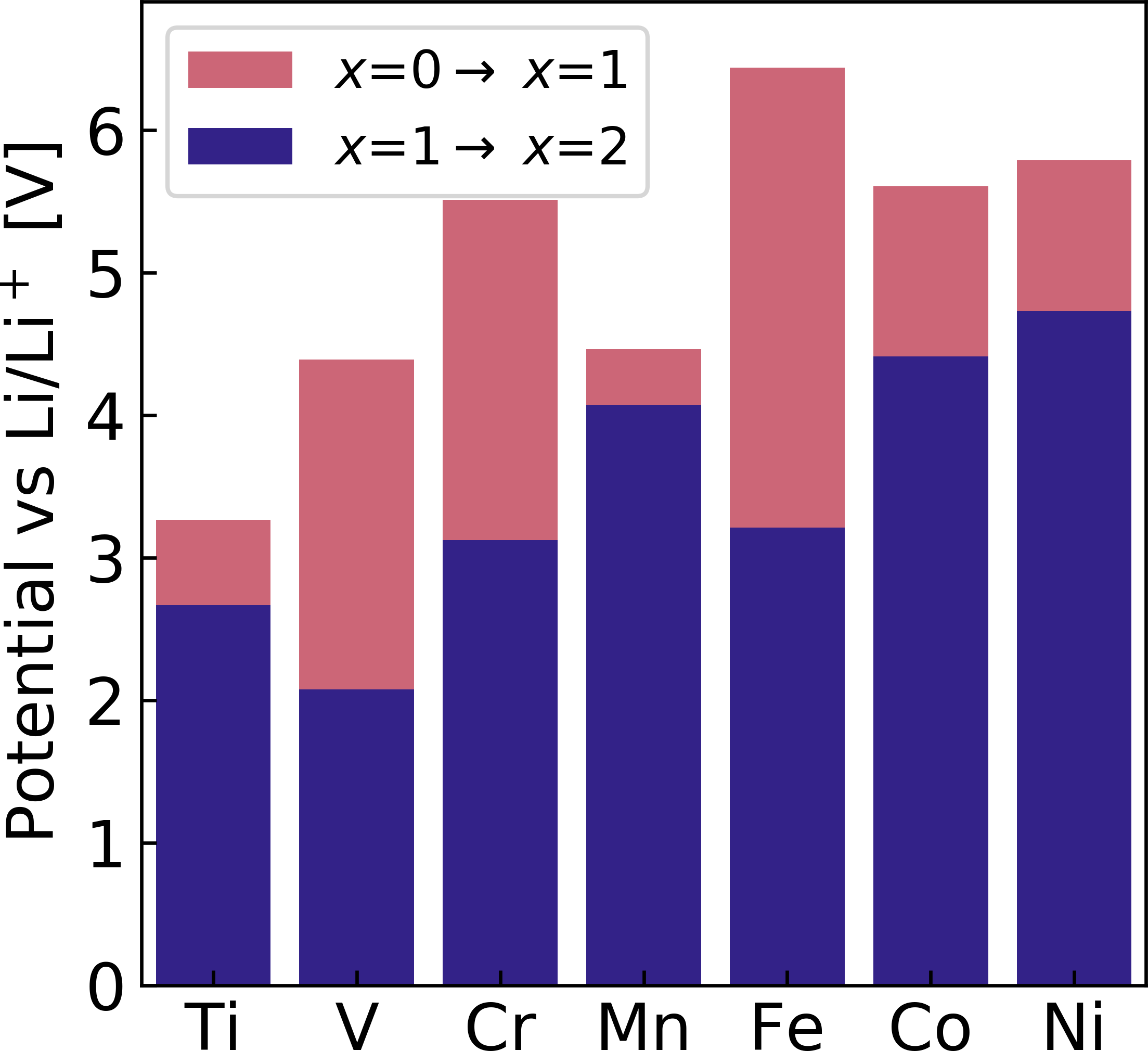}
    \caption{Theoretical voltages of the Li$_x$CaMF$\mathrm{_6}$ compounds for the two lithiation steps, $x$=0 to $x$=1 and $x$=1 to $x$=2.}
    \label{fig:theoretical_voltages}
\end{figure}

To assess the suitability of the colquiriit compounds as active cathode materials, we calculated their theoretical capacities and voltages. The capacity is given by the maximum ionic charge which can be stored in the compounds (here, two Li$\mathrm{^+}$ ions per formula unit) with respect to its mass (gravimetric capacity) or equilibrium volume (volumetric capacity). Considering the total theoretically accessible lithiation range ($x$=0 to $x$=2), the gravimetric (volumetric) capacitities lie in the range of 222.6 - 233.4 $\frac{mAh}{g}$ (240.9 - 263.3 $\frac{mAh}{cm^3}$), if referenced to the masses (equilibrium volumes) of the colquiriite compounds at $x$=2. In comparison, the corresponding capacities of Li$_x$CoO$\mathrm{_2}$ (LCO) are 273.8 $\frac{mAh}{g}$ (339.1 $\frac{mAh}{cm^3}$) in the range $x$=0 to $x$=1. However, the cyclability of LCO is known to be restricted to the range 0.5 $\leq x \leq$ 1, reducing the capacity accordingly to 137 $\frac{mAh}{g}$ \cite{deng_LCO_ref}.
The voltages can be determined from the total energies of the compounds \cite{urban_computational_2016}, which we obtained from our DFT+$U$ calculations. The voltages, referenced against the oxidation potential of metallic lithium, are displayed in Figure \ref{fig:theoretical_voltages} separately for the two lithiation steps. For all considered compounds, the voltage decreases from the first to the second lithiation step. This is in accordance with the finding, that the occupation of the WP 2c in the first step is energetically more favorable than the occupation of the WP 2a in the second step. Our calculated voltage for the intercalation step from CaCoF$\mathrm{_6}$ to LiCaCoF$\mathrm{_6}$ of 5.6 V is in good agreement with the value reported by Koyama \textit{et al.} (5.80 V) \cite{koyama_2000}. Koyama \textit{et al.} attributed this relatively high voltage, e.g. compared to the value of 4.2 V for Li$_x$CoO$\mathrm{_2}$, to the higher electronegativity of fluorine compared to oxygen. In line with this argument several considered compounds show higher voltages when compared to oxides utilizing the same redox couple (Ni$^{4+}$/Ni$^{3+}$: 4.3 V for Li$_x$NiO$\mathrm{_2}$, Mn$^{4+}$/Mn$^{3+}$: 3.4 V for Li$_x$NiO$\mathrm{_2}$ \cite{chakraborty_predicting_2018}). Only based on the calculated voltages, one could also imagine a cell with Li$_x$CaTiF$\mathrm{_6}$ as the anode cycled against Li$_x$CaFeF$\mathrm{_6}$ as the cathode, resulting in about 3 V in the Li range $x$=0 to $x$=1. The voltage would however drop to an unfavorable small value for the second lithiation step.


\section{Discussion}\label{sec:discussion}

\subsection{Choice of the $U$-parameter}\label{discussion:U_value_choosing}

The oxidation number of the electrochemically active \textit{3d} TM cation decreases when lithium ions are inserted into the colquiriite-type crystal structure and the cations have different ionic radii at different oxidation states. In a simple picture, in which the bond length between two ions is the sum of their ionic radii, the relative change in ionic radius, $\Delta r\mathrm{^{(rel.)}}$, can be related to the relative change of the MF$\mathrm{_6}$ octahedron volume, $\Delta V\mathrm{^{(rel.)}}$ (the derivation is presented in the Supplemental Material \cite{Supplemental_material}):

\begin{equation}\label{eq:octahedra_volume_Vs_Ionic_radii}
    \Delta V\mathrm{^{(rel.)}}=(\alpha \Delta r\mathrm{^{(rel.)}})^3 + 3 (\alpha \Delta r\mathrm{^{(rel.)}})^2+3 \alpha \Delta r\mathrm{^{(rel.)}}
\end{equation}

Here, $\alpha=\frac{r_0}{r_0+r_L}$ relates the change in ionic radius to the change in bond length, with $r_0$ and $r_L$ being the ionic radius of the M cation in the initial state and the ionic radius of the ligand, F$\mathrm{^-}$, respectively. The ionic radius of fluorine is 1.3 {\AA} for a three-fold coordinated ion in the charge state 1- and the typical values for the ionic radii of $3d$ TM lie in the range of 0.5 and 0.8 \AA.~Therefore $\alpha$ as well as $\Delta r\mathrm{^{(rel.)}}$ both are small ($<$1), and the change in volume of the octahedra can be approximated by the last, linear term of the right-hand side of Eq.\ (\ref{eq:octahedra_volume_Vs_Ionic_radii}).

\begin{figure}
    \centering
    \includegraphics[width=\columnwidth]{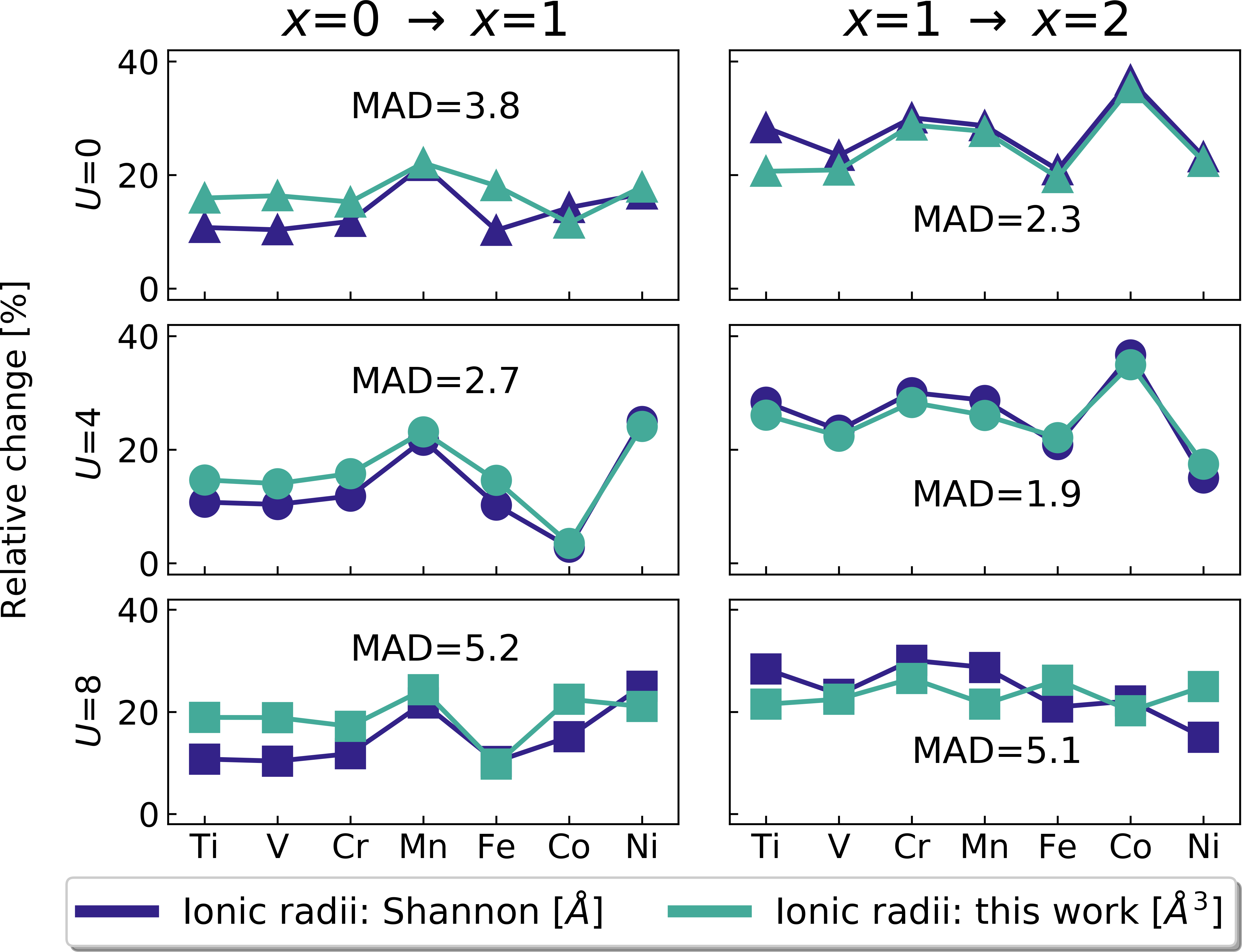}
    \caption{Comparison of the lithiation induced relative changes of ionic radii from Shannon and from the calculations performed in this work. The three panels on the left (right) display the results for the first (second) lithiation step. The $U$ correction on the M cations increases from top to bottom. Inserted into each panel is the mean absolute deviation (MAD) between the two sets of ionic radii changes.}
    \label{fig:ionic_radii_comparison}
\end{figure}

We choose the ionic radius to select an appropriate value for $U$, because of its experimental availability and the established linear correlation to the change in MF$\mathrm{_6}$ octahedron volume, an important quantity for the investigation of local and global volume changes due to Li insertion.
For all our considered cations in their oxidation states 4+, 3+, and 2+, the ionic radii derived from experimental data of mineral structures, with the different magnetic spin states taken into account, were compiled by Shannon \cite{Shannon_1976_ionic_radii}. We derived the corresponding ionic radii from the bond lengths in our calculated ground-state crystal structures.
Figure \ref{fig:ionic_radii_comparison} compares the change in Shannon ionic radii to the change of our calculated ionic radii for the two lithiation steps and the three considered $U$ values. Overall, the change in calculated radii fit well with the change in Shannon radii. For $U$=0 eV and $U$=4 eV, they even follow the same trend along the \textit{3d} series of the M cations, with a slight deviation only for M=Fe and M=Co in the first lithiation step for $U$=0 eV. For $U$=8 eV, the calculated values exhibit the strongest differences to the experimental ones, especially for the second lithiation step. In order to further quantify the agreement between the two sets, we calculated the mean absolute deviation (MAD) for each set of datapoints, which is displayed for each combination of lithiation step and $U$ value in Fig.\ \ref{fig:ionic_radii_comparison}.

For both lithiation steps the MAD is lowest for $U$=4 eV and for that value the overall trend along the series of M ions is well reproduced. This is in good agreement with other computational works on \textit{3d} TM fluorides: a value of 4 eV lays well in the range of previously used $U$ corrections (e.g., 5 eV for Fe in FeF$\mathrm{_3}$ \cite{Li_2010_DFTU}, 4 eV for Ni and 5 eV Co in AgMF$\mathrm{_4}$ (M=Ni, Co) \cite{Domanski_2020_DFTU}, 2.4 eV for Ti and 5.0 eV for Fe ions in Na$\mathrm{_2}$TiFeF$\mathrm{_7}$, and 4 eV for Mn in LiMn$\mathrm{_2}$O$\mathrm{_4}$ \cite{Ramogayana_2022_DFTU}).

\subsection{Changes in ionic radii along the \textit{3d} TM series}\label{sec:discussion_ionic_radii_curves}

As described in Sec.\ \ref{sec:results}, the magnitude of volume changes of the colquiriite-type crystal structure upon lithiation significantly depends on the species and size of the TM ion. Similar findings were reported by Ariyoshi \textit{et al.} for solid solutions of Li[Ni$\mathrm{_{1-x-y}}$Co$\mathrm{_{x}}$Mn$\mathrm{_{y}}$]O$\mathrm{_{2}}$ with $0\leq x \leq 1$ and $0\leq y\leq 1$ \cite{Ariyoshi_2019_ionic_radii}, where the ratio of the different TM species strongly influences the volume change in the lithiation process. Obviously, for both ligand elements, L=O and L=F, the volume changes of the ML$\mathrm{_{6}}$ octahedra is correlated with the changes of the ionic radii of the M ions. TM cations in lower oxidation states have larger ionic radii since the electrons are less strongly attracted by the smaller positive ionic charge. Therefore, the change in ionic radius is always positive with increasing Li concentration. The larger radii increase the bond lengths between the TM cations and the surrounding anions, and, accordingly, the corresponding volumes of the coordination octahedra. Therefore, in the following, we first discuss the changes in ionic radii along the \textit{3d} TM series according to the ligand field theory (LFT) before we connect the local volume changes of the octahedra surrounding the cations with the global volume changes of the crystal structures in Sec.\ \ref{sec:discussion_global_volume_change}.

According to the LFT, nine orbitals of the central TM cation can be involved in bonding in an octahedral coordination. For \textit{3d} TM, the involved atomic orbitals (AOs) are the \textit{3d}, \textit{4s} and \textit{4p} orbitals. The \textit{4s} and the \textit{4p} orbitals have the symmetry $a\mathrm{_{1g}}$ and $t\mathrm{_{1u}}$, respectively. The \textit{3d} orbitals are split into two symmetry groups, denoted $e\mathrm{_g}$ ($d_{x^2-y^2}$ and $d_{z^2}$) and $t\mathrm{_{2g}}$ ($d_{xy}$, $d_{xz}$, $d_{yz}$). In $\sigma$ bonding, six orbitals of the ligands (for fluorides: one \textit{2p} orbital per fluorine) form symmetry adapted linear combinations (SALCs), which can be split into three symmetry groups: $e\mathrm{_g}$, $a\mathrm{_{1g}}$ and $t\mathrm{_{1u}}$. In order to overlap (i.e., to form a $\sigma$-bond) the AOs of the central cation and the SALCs must have the same symmetry. With the exception of the $t\mathrm{_{2g}}$ orbitals, this is fulfilled for all AOs and SALCs.

The $t\mathrm{_{2g}}$ can form $\pi$ bonds with another set of SALCs of the ligands (formed by the other two \textit{2p} orbitals), which have the $t\mathrm{_{2g}}$ symmetry \cite{miessler_inorganic_2004_LFT}. In general, the $t\mathrm{_{2g}}$ orbitals could also interact with $t\mathrm{_{2g}}$ orbitals from other metal cations in the colquiriite structure. However, the MF$\mathrm{_6}$ octahedra are not directly connected to each other (i.e., no TM cation is the next nearest neigbhor to another TM cation) and are separated by more than 5.1 \AA~ in the $(0001)$ plane. Along the $[0001]$ direction they are separated by CaF$\mathrm{_6}$ octahedra, whose central cations have no \textit{d} orbitals to form $\pi$ bonds.

In summary, the orbitals with symmetry groups $e\mathrm{_g}$, $a\mathrm{_{1g}}$ and $t\mathrm{_{1u}}$ form $\sigma$ bonding and antibonding (indicated by an asterisk) orbitals. The orbitals with symmetry group $t\mathrm{_{2g}}$ form $\pi$ and $\pi^*$ orbitals. All the bonding orbitals are occupied by the number of electrons corresponding to the number of \textit{2p} electrons of the F$\mathrm{^-}$ ligands. The next orbitals that can be occupied are the antibonding $t\mathrm{_{2g}^*}$ and the $e\mathrm{_g^*}$ orbitals. \cite{Griffith_1957_LFT}. They are occupied by the number of electrons corresponding to the number of \textit{3d} electrons of the central TM. In general, antibonding orbitals increase the bond length. The $t\mathrm{_{2g}^*}$ form weaker $\pi$ bonds compared to the $e\mathrm{_g^*}$ $\sigma$ bonds. Therefore, the occupation of the former increases the M-F bond length to a lower extent. This has already been reported in other works, concerning \textit{3d} TM cathode materials \cite{Marianetti_2001,Zhao_2022}. As described in the context of Eq. (\ref{eq:octahedra_volume_Vs_Ionic_radii}), the M-F bond lengths are linearly related to the ionic radii of the M cations and to the MF$\mathrm{_6}$ octahedron volumes. 

\begin{figure}
    \centering
    \includegraphics[width=0.8\columnwidth]{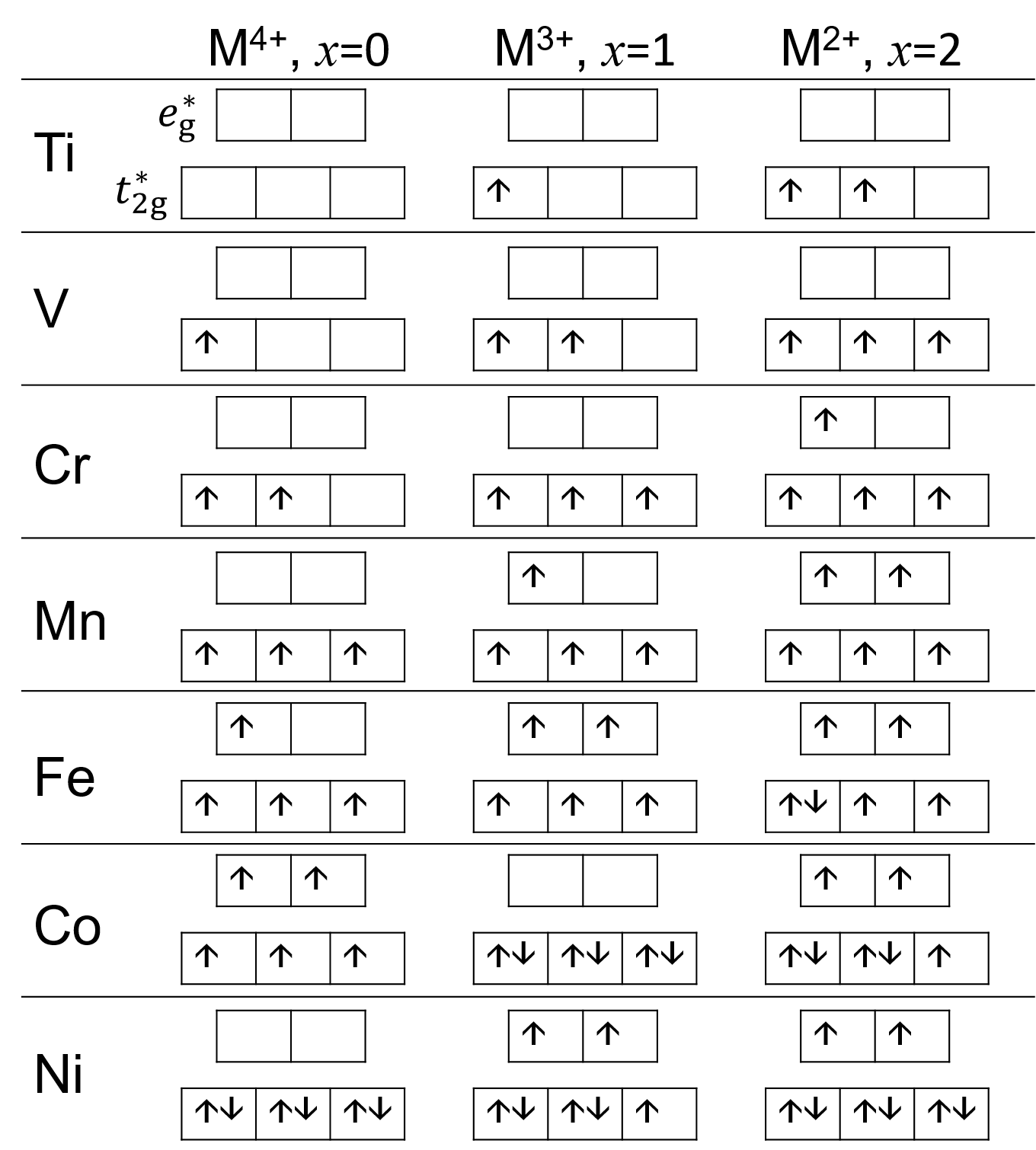}
    \caption{Formal \textit{3d} orbital occupations for the three oxidation states of the \textit{3d} TM cations M in their respective ground states corresponding to the different Li concentrations $x$. Along the series from Ti to Fe, the ground states are always magnetic HS states. The magnetic states of Co (Ni) follow the sequence HS$\rightarrow$LS$\rightarrow$HS (LS$\rightarrow$HS$\rightarrow$HS) with increasing $x$.}
    \label{fig:schematic_orbital_occuaption}
\end{figure}

We first discuss the relative change of the octahedron volumes during the first lithiation step for each M cation, as plotted in the left panel of Fig.\ \ref{fig:U4_all_octahedra_CellVolume}. Figure \ref{fig:schematic_orbital_occuaption} schematically depicts the formal spin configurations in the different oxidation states corresponding to the ground states obtained in the calculations (\textit{cf.} Tab.\ \ref{tab:delta_energies}). For Ti, V, and Cr, the additional electron of the first inserted Li atom occupies one of the previously empty $t\mathrm{_{2g}^*}$ orbitals. For Mn, however, one of the $e\mathrm{_g^*}$ orbitals becomes filled, which results in a considerable increase of the ionic radius. For Fe the change is again smaller, because in Fe$\mathrm{^{4+}}$ there was formally already one $e\mathrm{_g^*}$ orbital occupied and therefore the relative change due to the second $e\mathrm{_g^*}$ orbital becoming occupied is not as large.

\begin{figure}
    \centering
    \includegraphics[width=0.8\columnwidth]{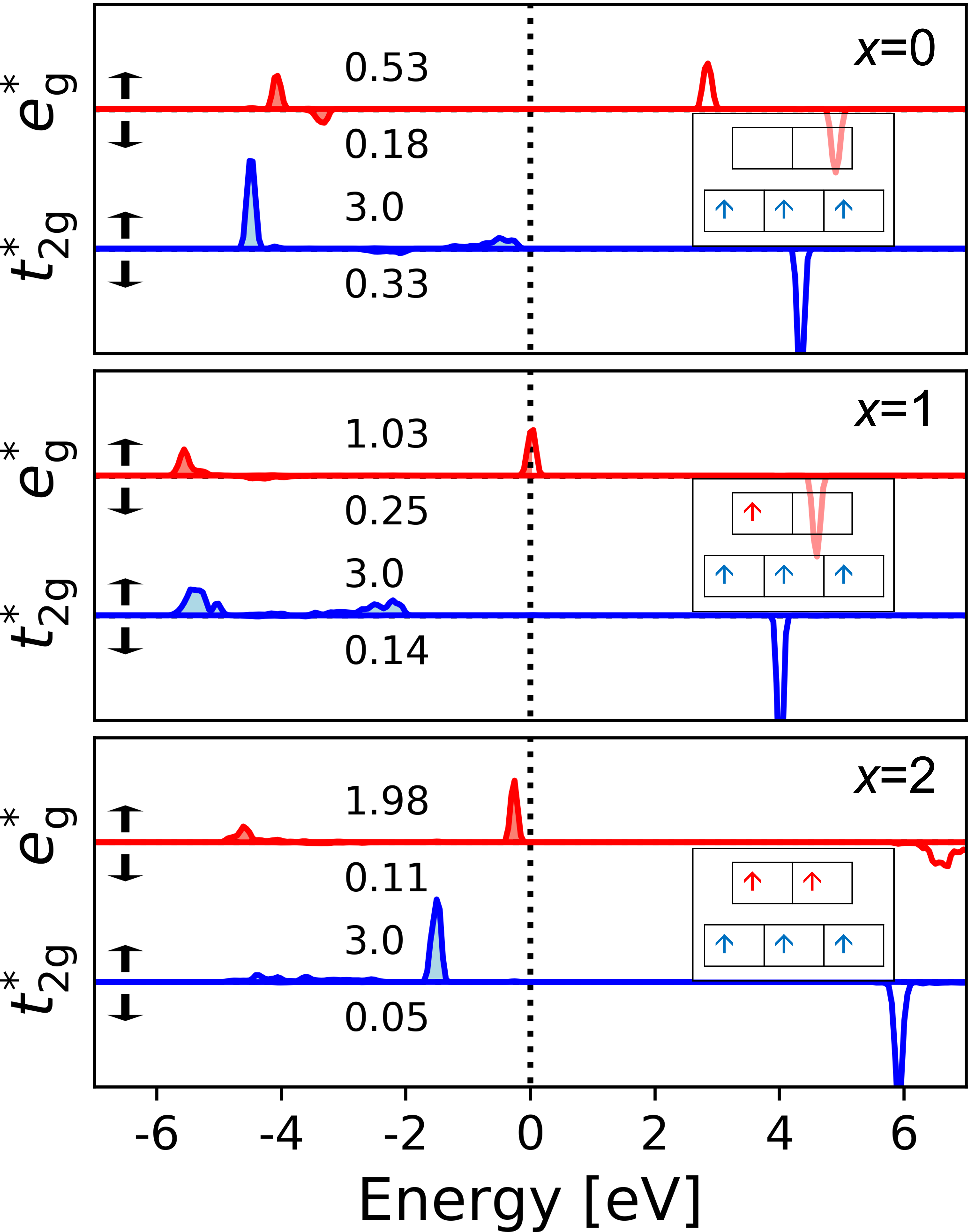}
    \caption{Site- and orbital-projected densities of states of the $t\mathrm{_{2g}^*}$ and $e\mathrm{_g^*}$ orbitals of Mn in Li$_x$CaMnF$\mathrm{_6}$, calculated with $U$=4 eV. The insets sketch the expected formal distributions of the \textit{3d} electrons in the high-spin states (ground states for Li$_x$CaMnF$\mathrm{_6}$) amongst the two orbital groups according to the LFT. The energies are referenced to the Fermi energy (dotted vertical lines). The numbers of electrons (or charges in units of the elementary charge) obtained by integrating over the occupied states in each orbital group and spin channel are written next to the corresponding curves.}
    \label{fig:licamnf_pdos}
\end{figure}

The small volume change of the CoF$\mathrm{_6}$ octahedron (1.6~\%) can be explained by the fact that for $x$=0 the HS state is energetically preferred, while for $x$=1 it is the LS state. This means that for $x$=0 both $e\mathrm{_g^*}$ orbitals are singly occupied, while for $x$=1 all three $t\mathrm{_{2g}^*}$ orbitals are doubly occupied by the 6 electrons. This relocation of the electrons nearly completely compensates the effect of the additional electron, which increases the ionic radius.

Ni (with formally 6 [Ni$\mathrm{^{4+}}$] or 7 [Ni$\mathrm{^{3+}}$] \textit{3d} electrons) exists in the LS state for $x$=0 and in the HS state for $x$=1. Therefore, at $x$=0 all $t\mathrm{_{2g}^*}$ orbitals are doubly occupied and at $x$=1, both $e\mathrm{_g^*}$ orbitals are singly occupied. As in the case of Mn, this also leads to a larger volume change of the M octahedron.

Corresponding arguments can be applied for the second lithiation step. The Cr and Co octahedra are expected to yield the largest positive changes, because for both systems previously empty $e\mathrm{_g^*}$ orbitals become occupied, which is reflected by the data points. The octahedron around Mn has a comparably large volume change, too, since here the second $e\mathrm{_g^*}$ orbital becomes occupied.

To strengthen the arguments made above based on the LFT, we relate them to the results of electronic structure calculations for site- and orbital-projected densities of states (PDOS) of the \textit{3d} orbitals of the TM ions. Figure \ref{fig:licamnf_pdos} illustrates, that for Li$_x$CaMnF$\mathrm{_6}$ the charges obtained by integrating the PDOS up to the energy of the highest occupied level are for both spin channels in good agreement with the formal charges expected from the LFT.
The largest deviation is obtained for the $e\mathrm{_g^*}$ orbitals at $x$=0 which show  non-zero occupations in both spin channels. However, they are expected to be empty according to the LFT. We can attribute this small mismatch to some electronic hybridization interaction of the well localized \textit{3d} orbitals with the more delocalized, formally empty \textit{4s} and \textit{4p} orbitals. Therefore, a small fraction of the electron density gets projected onto atom-centered partial waves of \textit{d}-orbital symmetry in the calculation and integration of the PDOS. 

In the first lithiation step, the state in the spin-up channel of the $e\mathrm{_g^*}$ orbital, which is unoccupied at $x$=0, shifts towards the Fermi level and becomes partially occupied at $x$=1. In the second lithiation step, this state is shifted below the Fermi level and is then completely occupied. This trend agrees with the LFT and supports the validity of the arguments made above.

\subsection{Effects of local volume changes on the volume change of the unit cell}\label{sec:discussion_global_volume_change}

To understand how the volume of the unit cell of the crystal is affected by the local volume changes of the fluorine coordinated octahedra around the different cations, we analyze the structural changes during lithiation with respect to the location of the octahedra and to their connections to each other in the colquiriite structure.

\begin{figure}
    \centering
    \includegraphics[width=\columnwidth]{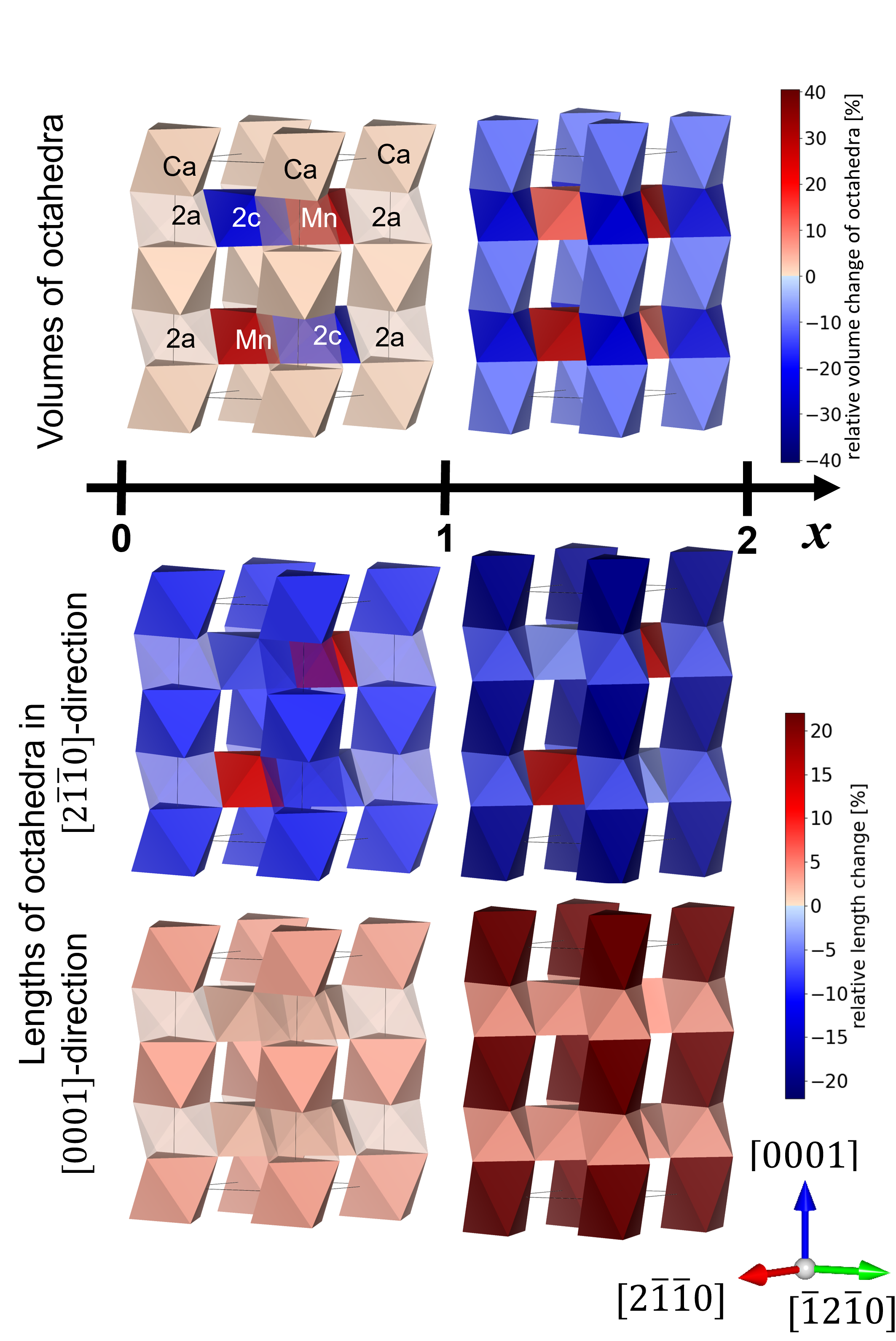}
    \caption{Change of structural parameters of Li$_x$CaMnF$\mathrm{_6}$ during the two lithiation steps. The upper two structures display the change in the octahedra volumes while the central and lower panels show the length changes of the octahedra along the lattice directions. Transparent octahedra represent octahedral sites that are not occupied. In the top left structure the Ca and Mn octahedra are labeled by the central cation and the Lithium octahedra are labeled with the respective Wyckoff position.}
    \label{fig:licamnf6_octaederchange}
\end{figure}

In Sec.\ \ref{sec:results_volchanges} we describe that the MF$\mathrm{_6}$ octahedra swell in the first lithiation step, whereas the F$\mathrm{_6}$ octahedra around the 2c sites, on which the Li ions are placed, shrink (\textit{cf.}\ Fig.\ \ref{fig:U4_all_octahedra_CellVolume}). This is shown schematically for Li$_x$CaMnF$\mathrm{_6}$ in Fig.\ \ref{fig:licamnf6_octaederchange} (upper left structure). MF$\mathrm{_6}$ and LiF$\mathrm{_6}$ (2c) octahedra are located next to each other in the (0001) plane and are connected by a common edge. This leads to an elongation of both types of octahedra in $[0001]$ direction by the same amount, while the edge lengths in the in-plane directions ($[2\Bar{1}\Bar{1}0]$ and $[\Bar{1}2\Bar{1}0]$) change oppositely for MF$\mathrm{_6}$ and LiF$\mathrm{_6}$ octahedra (lower left two structures in Fig.\ \ref{fig:licamnf6_octaederchange}). For the series of considered compounds, these compensating effects lead to changes of the unit cell lattice parameter $a$ between -4.2~\% and -1.0~\%, and $c$ between 0.9~\% and 2.6~\%. All values for the lattice parameters are summarized in the Supplemental Material \cite{Supplemental_material}. For M=Mn the deformations along the different crystallographic directions compensate each other in such a way that the volume change of the whole unit cell is less than 1~\%. Although the changes in lattice parameters $a$ and $c$ have different signs and magnitudes (-1.0~\% and 2.6~\%, respectively), which could lead to stresses in a confined single crystal material, we would expect an averaging out of the different expansions/contractions to some extent in a polycrystalline material and at the corresponding grain boundaries.

The volume change for Li$_x$CaCoF$\mathrm{_6}$ of -7.4~\% that we find deviates considerably from the value of -0.4~\% reported by Koyama \textit{et al.} \cite{koyama_2000}. The difference may at least partially be attributed to the inclusion of spin-polarization and the application of a $U$ correction in our work, which are both essential to get different magnetic states related to different oxidation states of the electroactive M cations, as discussed before. Spin-polarization and electron-correlation effects were not yet taken into account in the work of Koyama.

For the lithiation step from $x$=1 to $x$=2 the unit-cell-volume changes are positive for all the considered compounds and larger in magnitude, although the MF$\mathrm{_6}$ octahedra and the LiF$\mathrm{_6}$ octahedra behave similarly as for $x$=0 to $x$=1. The former swell and the latter shrink. This again leads to a compensation effect resulting in changes of the lattice parameter $a$ between -0.6 and +0.6~\% for the set of considered compounds in the second lithiation step.

However, the increase of the lattice parameter $c$ is more pronounced and leads in total to volume changes of the unit cells between 10~\% and 15~\% as displayed in Fig.\ \ref{fig:U4_all_octahedra_CellVolume} (right panel). This can be explained as follows:\\
First, one can see in Fig.\ \ref{fig:licamnf6_octaederchange} that the MF$\mathrm{_6}$ octahedra expand due to the oxidation in all three lattice directions $[2\Bar{1}\Bar{1}0]$, $[\Bar{1}2\Bar{1}0]$ and $[0001]$. The elongation in $[0001]$ direction is transferred with the same extent to the neighboring LiF$\mathrm{_6}$ (2c) and LiF$\mathrm{_6}$ (2a) octahedra in the (0001) plane due to their edge connections (Fig.\ \ref{fig:licamnf6_octaederchange}, bottom right structure). The occupation of the 2a positions by Li leads to a reduced electrostatic repulsion among the six surrounding F anions and an overall shrinkage of the LiF$\mathrm{_6}$ (2a) octahedra. But since these increase in $[0001]$ direction, the shrinkage in volume can only be realized by a reduction of the edge lengths in the $[2\Bar{1}\Bar{1}0]$ and $[\Bar{1}2\Bar{1}0]$ lattice directions. This is possible because there are no constraints imposed by the MF$\mathrm{_6}$ octahedra for this structural adaption.

The top and bottom faces of the LiF$\mathrm{_6}$ (2a) octahedra in $[0001]$ direction are shared with the CaF$\mathrm{_6}$ octahedra, which accordingly also shrink in the $[2\Bar{1}\Bar{1}0]$ and $[\Bar{1}2\Bar{1}0]$ directions together with the LiF$\mathrm{_6}$ (2a) octahedra.

Since the Ca cations do not change their oxidation state, no change in the average Ca-F bond length should occur. The trade-off between the reduced areas of the top and bottom faces of the Ca octahedra and the resistance against shorter Ca-F bonds results in an elongation of the CaF$_6$ octahedra along the $[0001]$ lattice direction. This can be accomplished by the colquiriite crystal structure up to a certain extent since the CaF$\mathrm{_6}$ octahedra are not constrained by other octahedra in this direction. A complete compensation, however, cannot be achieved, leading to a volume reduction for all CaF$\mathrm{_6}$ octahedra (see Fig.\ \ref{fig:U4_all_octahedra_CellVolume}).

The elongation of octahedra along the $[0001]$ lattice direction yields changes in the $c$ lattice parameter of the unit cell between 11.2~\% and 11.8~\% for the different M occupations and therefore an increase of the unit-cell volumes. The variation of the values for $c$ is less than for $a$ and also less than the change in $c$ during the first lithiation step. In the second lithiation step the influence of the type of M cation on the volume change appears to be less pronounced than in the first lithiation step.

The theoretical calculated gravimetric capacity of 119.8 $\frac{mAh}{g}$ of Li$_x$CaFeF$\mathrm{_6}$ for the second lithiation step ($x$=1 to $x$=2) is in good agreement with the measured capacity of 112 $\frac{mAh}{g}$ for the lithiation range from 1 $\leq x \leq$ 1.8 \cite{de_biasi_licafef_2017}. Also, our calculated value for the equilibrum voltage of 3.2 V lies well in the range of the measured voltages of 2.0--4.5 V. However the calculated volume changes do not agree with the experimentally measured volume changes of $\Delta V_0^{\rm (rel.)} \leq$ 0.5~\%. In line with the experimental results, we obtained Fe$\mathrm{^{3+}}$ octahedra (9.9 \AA$^3$) with smaller volumes as the LiF$\mathrm{_6}$ (2a) octahedra (10.8 \AA$^3$), but in contrast to the experimental results we found at $x$=2 larger octahedra around the Fe$\mathrm{^{2+}}$ (10.3 \AA$^3$) than around the now occupied Li ions (11.7 \AA$^3$), \textit{cf.}\ Fig.\ \ref{fig:absolute_octahedra_volumes}. In addition, the stretching of the Ca octahedra during the second lithiation step was not reported by de Biasi \textit{et al.} Note that the cell volumes (and therefore the relative volume changes) only show a weak dependence with the used $U$ correction, as long as the same magnetic states are stable (all equilibrium volumes are given in the Supplemental Material \cite{Supplemental_material}). For Li$_x$CaFeF$\mathrm{_6}$ the maximum difference in cell volumes for the same Li concentration amounts to 3.9 \AA$^3$~for $x$=1 with volumes between 229 \AA$^3$~ ($U$=8 eV) and 233 \AA$^3$~ ($U$=0 eV). Therefore the deviation to the experimental results cannot be attributed to an inappropriate choice of the $U$ correction value. This leads us to the conclusion that the experimentally observed ZS behavior in Li$_x$CaFeF$\mathrm{_6}$ cannot fully be understood as a single-crystal phenomenon. Other possible effects that result in a measured low volume change may be caused by Li ions at grain boundaries, other charged point defects (e.g., cation vacancies), or amorphous surface and interface layers.


\section{Conclusions} \label{sec:conclusion}

We present a systematic DFT+$U$ investigation of volume changes in colquiriite-type crystalline  compounds Li$_x$CaMF$\mathrm{_6}$ with varying Li concentration. Considering on the M positions different cations along the series of \textit{3d} transition-metal elements of the periodic table, we explored the influence of a Hubbard $U$ correction on the magnetic ground states of the compounds. By comparing ionic radii obtained from those ground state structures to Shannon’s ionic radii we determined an appropriate $U$ value of 4 eV. Trends in local and global structural changes along the \textit{3d} series of the TM elements are elucidated by relating the electronic structure results to predictions of the LFT. We explain the connection of volume changes of the fluorine coordinated octahedra surrounding the cations to the change of the unit cell volume. All the different octahedra have partly constructive, partly compensating influences on the volume change of the crystal unit cell.\\
Due to the large volume increase of the MF$\mathrm{_6}$ octahedra from M$\mathrm{^{3+}}$ to M$\mathrm{^{2+}}$ and the stretching of the Ca octahedra in the second lithiation step, the unit cell volume increases from $x$=1 to $x$=2  by approximately 10~\% for all structures. This is not in line with experimental results for Li$_x$CaFeF$\mathrm{_6}$. This discrepancy may originate from extended defects in polycrystalline microstructures, such as grain boundaries, which are not covered in the single-crystal model of this work.\\
From $x$=0 to $x$=1 the different volume changes of the MF$\mathrm{_6}$ octahedra follow the trend of the change in ionic radii and determine the magnitude and sign of the change of the unit-cell volume. The trend along the \textit{3d} TM series can be explained by the successive occupations of the two orbital groups $t\mathrm{_{2g}^*}$ and $e\mathrm{_g^*}$ by the additional electrons of the inserted Li atoms.\\
We found in our work that Li$_x$CaMnF$\mathrm{_6}$ is a promising ZS candidate for a LIB cathode material with a voltage and a theoretical capacity comparable to capacities of common cathode materials \cite{NITTA2015252}. To the best of our knowledge, this compound has only been synthesized for $x$=0 in another structure type, namely one of binary fluorides, VF$\mathrm{_3}$, \cite{Hoppe_1957_CaMnF6} and its stability in the colquiriite structure as well as its suitability as cathode material has not yet been experimentally investigated.

\acknowledgments
This work was funded by the German Research Foundation (DFG, Grant No. EL 155/29-1). The authors acknowledge support by the state of Baden-Württemberg through bwHPC and the German Research Foundation (DFG) through grant no INST 40/575-1 FUGG (JUSTUS 2 cluster). Crystallographic drawings were created using the software VESTA \cite{Momma_2011_VESTA}.\\

\bibliography{references}

\end{document}


\title{Supplemental Material: \textit{First-principles} analysis of the interplay between electronic structure and volume change in colquiriite compounds during Li intercalation}

\author{A. F. Baumann}
 \affiliation{University of Freiburg, Freiburg Materials Research Center (FMF), Stefan-Meier-Straße 21, 79104 Freiburg, Germany}
 
\author{D. Mutter}%
 \affiliation{Fraunhofer IWM, Wöhlerstraße 11, 79108 Freiburg, Germany}

\author{D. F. Urban}
\affiliation{University of Freiburg, Freiburg Materials Research Center (FMF), Stefan-Meier-Straße 21, 79104 Freiburg, Germany}
\affiliation{Fraunhofer IWM, Wöhlerstraße 11, 79108 Freiburg, Germany}

\author{C. Elsässer}
\affiliation{University of Freiburg, Freiburg Materials Research Center (FMF), Stefan-Meier-Straße 21, 79104 Freiburg, Germany}
\affiliation{Fraunhofer IWM, Wöhlerstraße 11, 79108 Freiburg, Germany}

\maketitle

\tableofcontents

\subsection{Magnetic moments of the TM ions}
\noindent Table \ref{tab:ground_state_spins} lists the magnetic moments of the TM ions for the colquiriite compounds Li$_x$CaMF$\mathrm{_6}$, M=Ti, V, Cr, Mn, Fe, Co and Ni, $x$=0, 1, 2, and for the applied $U$ corrections of $U$=0 eV, 4 eV and 8 eV.\\

\begin{ruledtabular}
\begin{table}[ht]
\centering
\begin{tabularx}{0.5\textwidth}{l|cccccccc}
\textrm{$x$=0; M$^\mathrm{4+}$}& \textrm{Ti $d^{0}$ }   & \textrm{V $d^{1}$ }    & \textrm{Cr $d^{2}$  }  & \textrm{Mn $d^{3}$  }  & \textrm{Fe $d^{4}$ }   & \textrm{Co $d^{5}$ }   & \textrm{Ni $d^{6}$}  \\ \hline
$U$=0  & 0.0 & 1.0 & 1.9 & 2.7 & 3.3 & 0.8 & 0.0 \\
$U$=4  & 0.0 & 1.0 & 2.1 & 3.0 & 3.8 & 3.5 & 0.0 \\
$U$=8  & 0.0 & 1.1 & 2.3 & 3.4 & 3.8 & 3.9 & 0.0\\\\
\textrm{$x$=1; M$^\mathrm{3+}$}& \textrm{Ti $d^{1}$ }   & \textrm{V $d^{2}$ }    & \textrm{Cr $d^{3}$  }  & \textrm{Mn $d^{4}$  }  & \textrm{Fe $d^{5}$ }   & \textrm{Co $d^{6}$ }   & \textrm{Ni $d^{7}$}  \\ \hline
$U$=0  & 1.0 & 1.9 & 2.8 & 3.7 & 4.2 & 0.0 & 0.8\\
$U$=4  & 1.0 & 2.0 & 3.0 & 4.0 & 4.4 & 0.0 & 2.2\\
$U$=8  & 1.1 & 2.1 & 3.1 & 4.3 & 4.8 & 3.4 & 2.5\\\\
\textrm{$x$=2; M$^\mathrm{2+}$}& \textrm{Ti $d^{2}$ }   & \textrm{V $d^{3}$ }    & \textrm{Cr $d^{4}$  }  & \textrm{Mn $d^{5}$  }  & \textrm{Fe $d^{6}$ }   & \textrm{Co $d^{7}$ }   & \textrm{Ni $d^{8}$}  \\ \hline
$U$=0  & 1.8 & 2.8 & 3.7 & 4.5 & 3.7 & 2.7 & 1.7 \\
$U$=4  & 1.9 & 2.9 & 3.9 & 4.7 & 3.8 & 2.8 & 1.8\\
$U$=8  & 2.0 & 3.0 & 4.1 & 4.8 & 4.0 & 2.9 & 1.9\\\\
\end{tabularx}
\caption{Magnetic moments of the TM ions in the ground states of Li$_x$CaMF$\mathrm{_6}$ at the three Li concentrations $x$=0, 1, 2 and for the $U$ corrections, $U$=0 eV, 4 eV, 8 eV. All values are given in  units of $\mu_{\mathrm{B}}$}
\label{tab:ground_state_spins}
\end{table}
\end{ruledtabular}

\subsection{Equilibrium volumes and lattice constants}

\noindent Table \ref{tab:ground_state_volumes_U4} lists the ground state volumes, $V_0$, for the studied colquiriite compounds Li$_x$CaMF$\mathrm{_6}$ for the applied $U$ corrections of $U$=0 eV, 4 eV and 8 eV. Table \ref{tab:ground_state_lattice} lists the values of the corresponding lattice parameters $a$ and $c$.

\begin{ruledtabular}
\begin{table}[ht]
\centering
\begin{tabular}{lccccccc} 
\textrm{$x$=0; M$^\mathrm{4+}$}& \textrm{Ti $d^{0}$ }   & \textrm{V $d^{1}$ }    & \textrm{Cr $d^{2}$  }  & \textrm{Mn $d^{3}$  }  & \textrm{Fe $d^{4}$ }   & \textrm{Co $d^{5}$ }   & \textrm{Ni $d^{6}$}  \\ \hline
$U$=0  & 250.80 & 240.53 & 234.76 & 231.04 & 236.79 & 223.97 & 223.43\\
$U$=4  & 251.53 & 242.33 & 235.65 & 232.10 & 236.50 & 238.42 & 221.54\\
$U$=8  & 253.79 & 245.15 & 236.87 & 234.08 & 237.50 & 229.55 & 218.79\\
\\
\textrm{$x$=1; M$^\mathrm{3+}$}& \textrm{Ti $d^{1}$ }   & \textrm{V $d^{2}$ }    & \textrm{Cr $d^{3}$  }  & \textrm{Mn $d^{4}$  }  & \textrm{Fe $d^{5}$ }   & \textrm{Co $d^{6}$ }   & \textrm{Ni $d^{7}$}  \\ \hline
$U$=0  & 244.56 & 233.08 & 229.54 & 231.77 & 233.14 & 220.73 & 224.03\\
$U$=4  & 243.01 & 236.77 & 232.06 & 233.47 & 232.24 & 220.81 & 226.33\\
$U$=8  & 247.40 & 240.17 & 234.20 & 235.95 & 229.23 & 228.24 & 222.09\\
\\
\textrm{$x$=2; M$^\mathrm{2+}$}& \textrm{Ti $d^{2}$ }   & \textrm{V $d^{3}$ }    & \textrm{Cr $d^{4}$  }  & \textrm{Mn $d^{5}$  }  & \textrm{Fe $d^{6}$ }   & \textrm{Co $d^{7}$ }   & \textrm{Ni $d^{8}$}  \\ \hline
$U$=0  & 260.38 & 254.62 & 257.91 & 260.72 & 254.63 & 249.83 & 245.82\\
$U$=4  & 268.01 & 260.61 & 260.29 & 261.72 & 255.00 & 250.13 & 245.19\\
$U$=8  & 274.39 & 265.59 & 261.67 & 261.94 & 254.04 & 249.42 & 243.67\\
\end{tabular}
\caption{Equilibrium volumes $V\mathrm{_0}$ of the ground states for the Li$_x$CaMF$\mathrm{_6}$ structures at the three Li concentrations $x$={0, 1, 2} and for the $U$ corrections, $U$=0 eV, 4 eV and 8 eV. All values are given in \AA$^3$}
\label{tab:ground_state_volumes_U4}
\end{table}
\end{ruledtabular}

\begin{ruledtabular}
\begin{table}[ht]
\centering
\begin{tabular}{l|ccccccc}
\multicolumn{8}{c}{a lattice parameter}\\
\textrm{$x$=0; M$^\mathrm{4+}$}& \textrm{Ti $d^{0}$ }   & \textrm{V $d^{1}$ }    & \textrm{Cr $d^{2}$  }  & \textrm{Mn $d^{3}$  }  & \textrm{Fe $d^{4}$ }   & \textrm{Co $d^{5}$ }   & \textrm{Ni $d^{6}$}  \\ 
\hline
$U$=0  & 5.41 & 5.32 & 5.28 & 5.25 & 5.29 & 5.20 & 5.19\\
$U$=4  & 5.40 & 5.33 & 5.28 & 5.25 & 5.28 & 5.31 & 5.18\\
$U$=8  & 5.40 & 5.31 & 5.29 & 5.27 & 5.28 & 5.24 & 5.16\\
\\
\textrm{$x$=1; M$^\mathrm{3+}$}& \textrm{Ti $d^{1}$ }   & \textrm{V $d^{2}$ }    & \textrm{Cr $d^{3}$  }  & \textrm{Mn $d^{4}$  }  & \textrm{Fe $d^{5}$ }   & \textrm{Co $d^{6}$ }   & \textrm{Ni $d^{7}$}  \\ 
\hline
$U$=0  & 5.24 & 5.20 & 5.17 & 5.19 & 5.20 & 5.09 & 5.11\\
$U$=4  & 5.26 & 5.24 & 5.19 & 5.20 & 5.19 & 5.09 & 5.15\\
$U$=8  & 5.34 & 5.28 & 5.21 & 5.22 & 5.17 & 5.14 & 5.11\\
\\
\textrm{$x$=2; M$^\mathrm{2+}$}& \textrm{Ti $d^{2}$ }   & \textrm{V $d^{3}$ }    & \textrm{Cr $d^{4}$  }  & \textrm{Mn $d^{5}$  }  & \textrm{Fe $d^{6}$ }   & \textrm{Co $d^{7}$ }   & \textrm{Ni $d^{8}$}  \\ 
\hline
$U$=0  & 5.21 & 5.16 & 5.19 & 5.21 & 5.16 & 5.12 & 5.07\\
$U$=4  & 5.34 & 5.21 & 5.21 & 5.23 & 5.16 & 5.12 & 5.07\\
$U$=8  & 5.35 & 5.26 & 5.22 & 5.23 & 5.16 & 5.12 & 5.06\\
\\
\multicolumn{8}{c}{c lattice parameter}\\
\textrm{$x$=0; M$^\mathrm{4+}$}& \textrm{Ti $d^{0}$ }   & \textrm{V $d^{1}$ }    & \textrm{Cr $d^{2}$  }  & \textrm{Mn $d^{3}$  }  & \textrm{Fe $d^{4}$ }   & \textrm{Co $d^{5}$ }   & \textrm{Ni $d^{6}$}  \\ 
\hline
$U$=0  &  9.90 & 9.80 & 9.73 & 9.69 & 9.77 & 9.57 & 9.58\\
$U$=4  &  9.98 & 9.88 & 9.77 & 9.71 & 9.80 & 9.75 & 9.53\\
$U$=8  & 10.05 & 9.95 & 9.79 & 9.74 & 9.80 & 9.64 & 9.48\\
\\
\textrm{$x$=1; M$^\mathrm{3+}$}& \textrm{Ti $d^{1}$ }   & \textrm{V $d^{2}$ }    & \textrm{Cr $d^{3}$  }  & \textrm{Mn $d^{4}$  }  & \textrm{Fe $d^{5}$ }   & \textrm{Co $d^{6}$ }   & \textrm{Ni $d^{7}$}  \\ 
\hline
$U$=0  & 10.00 & 9.95 & 9.91 & 9.95 & 9.96 & 9.85 & 9.89\\
$U$=4  & 10.01 & 9.95 & 9.93 & 9.96 & 9.94 & 9.84 & 9.86\\
$U$=8  & 10.03 & 9.96 & 9.95 & 9.98 & 9.90 & 9.97 & 9.81\\
\\
\textrm{$x$=2; M$^\mathrm{2+}$}& \textrm{Ti $d^{2}$ }   & \textrm{V $d^{3}$ }    & \textrm{Cr $d^{4}$  }  & \textrm{Mn $d^{5}$  }  & \textrm{Fe $d^{6}$ }   & \textrm{Co $d^{7}$ }   & \textrm{Ni $d^{8}$}  \\ 
\hline
$U$=0  & 11.07 & 11.05 & 11.06 & 11.07 & 11.06 & 11.02 & 11.03\\
$U$=4  & 11.14 & 11.07 & 11.07 & 11.06 & 11.05 & 11.01 & 11.01\\
$U$=8  & 11.08 & 11.08 & 11.07 & 11.06 & 11.04 & 10.99 & 10.98\\
\end{tabular}
\caption{Lattice parameters $a$ and $c$ of the ground states for the Li$_x$CaMF$\mathrm{_6}$ structures at the three Li concentrations $x$={0, 1, 2} and for the $U$ corrections, $U$=0 eV, 4 eV and 8 eV. All values are given in \AA}
\label{tab:ground_state_lattice}
\end{table}
\end{ruledtabular}
\clearpage

\subsection{From the change in ionic radius to the change in octahedron volume}\label{app:ionic radius}

\begin{figure}[h]
    \centering
    \includegraphics[width=0.5\linewidth]{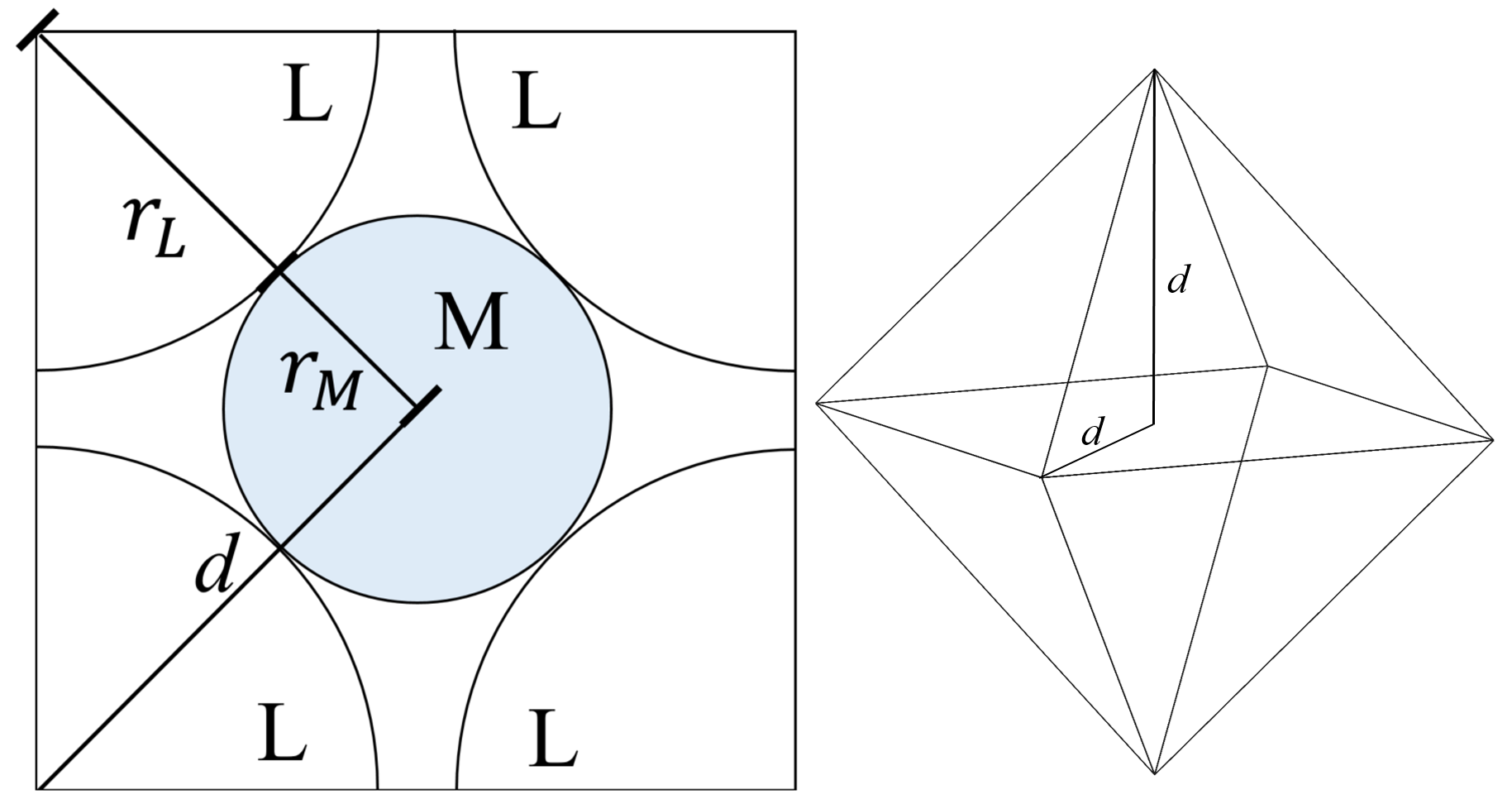}
    \caption{Left: Base area of the ML$\mathrm{_6}$ octahedron, where the ions are represented by hard spheres to visualize a definition of ionic radii. $r_M$, $r_L$, and $d$ are the ionic radii and the bond length, respectively. Right: Sketch of a regular octahedron, where all M-L bond lengths $d$ are equal.}
    \label{fig:appendix_schematics}
\end{figure}

\noindent The change in volume of a general octahedron can be related to the change in the ionic radius of the central cation. In Figure \ref{fig:appendix_schematics} (left) ionic radii are schematically used to define the bond length $d$ between the central cation M and the ligands, L (e.g., the fluorine anions):

\begin{equation}
    d=r_L+r_M,
\label{eq:bond_length_def}
\end{equation}

\noindent where $r_L$ and $r_M$ are the ionic radii of the ligand and the M cation, respectively.\\

\noindent We define:

\begin{equation}
    \Delta d\mathrm{^{(rel.)}}=\frac{d_1-d_0}{d_0},~~\Delta r\mathrm{^{(rel.)}}=\frac{r_1-r_0}{r_0},
\end{equation}

\noindent where $\Delta d\mathrm{^{(rel.)}}$ and $\Delta r\mathrm{^{(rel.)}}$ are the relative changes in bond length and ionic radius, respectively. The subscripts 0 and 1 correspond to two different states of the quantity (e.g., different oxidation states of the central cation and therefore different bond lengths/ionic radii), respectively.\\

\noindent Using Equation (\ref{eq:bond_length_def}), we obtain:

\begin{equation}
    \Delta d\mathrm{^{(rel.)}}=\frac{d_1-d_0}{d_0}=\frac{(r_1+r_L)-(r_0+r_L)}{(r_0+r_L)} = \frac{r_0}{r_0+r_L} \Delta r\mathrm{^{(rel.)}} = \alpha \Delta r\mathrm{^{(rel.)}},
\end{equation}

\noindent where the factor $\alpha=\frac{r_0}{r_0+r_L}$ relates the relative change in ionic radius to the relative change in bond length. The right panel of Figure \ref{fig:appendix_schematics} shows a sketch of a regular octahedron, in which all bond lengths $d$ are equal. The volume of such an octahedron can be calculated as the sum of the volume of two square pyramids:

\begin{equation}\label{eq:octahedron_volume}
    V=2 \frac{1}{3} A h=\frac{4}{3} d^3
\end{equation}

\noindent with the base area $A=\frac{(2d)^2}{2}$ and the height $h=d$. We can now express the relative change in the octahedron volume as a function of the relative change in bond length $\Delta d\mathrm{^{(rel.)}}$ and finally as a function of change in the ionic radius $\Delta r\mathrm{^{(rel.)}}$:

\begin{equation}\label{eq:octahedra_volume_Vs_Ionic_radii}
    \Delta V\mathrm{^{(rel.)}}=\frac{V_1-V_0}{V_0}=\frac{d_1^3-d_0^3}{d_0^3}=\left(\Delta d\mathrm{^{(rel.)}}\right)^3 +3\left( \Delta d\mathrm{^{(rel.)}}\right)^2 + 3 \Delta d\mathrm{^{(rel.)}}=(\alpha \Delta r\mathrm{^{(rel.)}})^3 + 3 (\alpha \Delta r\mathrm{^{(rel.)}})^2+3 \alpha \Delta r\mathrm{^{(rel.)}}
\end{equation}

\clearpage